\documentclass{article}
\usepackage{authblk}
\usepackage[utf8]{inputenc}
\usepackage[T1]{fontenc}
\usepackage[a4paper]{geometry}
\usepackage{graphicx}
\usepackage[textwidth=8em,textsize=small]{todonotes}
\usepackage{amsmath,amsthm,amssymb}
\usepackage{natbib}
\usepackage{hyperref}
\usepackage[nameinlink]{cleveref}
\usepackage{subcaption}
\usepackage{bbm}
\usepackage{amsfonts}
\usepackage{mathtools}
\usepackage[titletoc,toc]{appendix}
\usepackage{url}
\usepackage{algorithm}
\usepackage{algpseudocode}
\usepackage{multirow}
\usepackage{booktabs}
\usepackage[normalem]{ulem}
\usepackage{threeparttable}
\usepackage{enumitem}
\usepackage{array}
\usepackage{setspace}
\usepackage{orcidlink}

\definecolor{darkblue}{HTML}{0c7dbb}
\definecolor{darkgreen}{rgb}{0,0.48,0.65}
\hypersetup{
	colorlinks = true,
	citecolor=darkgreen,
	filecolor=black,
	linkcolor=darkblue,
	urlcolor=blue
}
\providecommand{\keywords}[1]
{
	\small
	\textbf{\textit{Keywords---}} #1
}

\DeclareMathOperator*{\argmin}{arg\,min}

\Crefname{algocf}{Algorithm}{Algorithms}
\Crefname{lemma}{Lemma}{Lemmas}
\title{Quantile Function-Based Models for Neuroimaging Classification Using Wasserstein Regression}
\author{Jie Li\textsuperscript{1,2}\orcidlink{0000-0001-8353-1322}\thanks{Corresponding author: Jie Li, School of Mathematics, Statistics and Actuarial Science, University of Kent, Canterbury, CT2 7NF UK.\ \textbf{Email}:~\href{jl725@kent.ac.uk}{jl725@kent.ac.uk}}}
\author[2,3]{Gary Green}
\author[1]{Jian Zhang}
\affil[1]{School of Mathematics, Statistics and Actuarial Science, University of Kent, Canterbury, CT2 7NF, UK}
\affil[2]{Innovision IP Ltd., 50 Seymour Street, London, England, W1H 7JG}
\affil[3]{York Neuroimaging Centre, University of York, Innovation Way, York, YO10 5NY, UK}

\begin{document}
\maketitle
\begin{abstract}
    We propose a novel quantile function-based approach for neuroimaging classification using Wasserstein-Fréchet regression, specifically applied to the detection of mild traumatic brain injury (mTBI) based on the MEG and MRI data. Conventional neuroimaging classification methods for mTBI detection typically extract summary statistics from brain signals across the different epochs, which may result in the loss of important distributional information, such as variance, skewness, kurtosis, etc. Our approach treats complete probability density functions of epoch space results as functional response variables within a Wasserstein-Fréchet regression framework, thereby preserving the full distributional characteristics of epoch results from $L_{1}$ minimum norm solutions. The global Wasserstein-Fréchet regression model incorporating covariates (age and gender) allows us to directly compare the distributional patterns between healthy control subjects and mTBI patients. The classification procedure computes Wasserstein distances between estimated quantile functions from control and patient groups, respectively. These distances are then used as the basis for diagnostic decisions. This framework offers a statistically principled approach to improving diagnostic accuracy in mTBI detection. In practical applications, the test accuracy on unseen data from Innovision IP's  dataset achieves up to 98\%.
\end{abstract}
\section{Introduction}\label{sec:introduction}
Mild traumatic brain injury (mTBI) affects millions of people annually and can lead to persistent cognitive and affective symptoms despite often unremarkable findings on conventional structural neuroimaging~\citep{lewineNeuromagneticAssessmentPathophysiologic1999,lewineObjectiveDocumentationTraumatic2007,huangSinglesubjectbasedWholebrainMEG2014,allenMagnetoencephalographyAbnormalitiesAdult2021}.
This diagnostic gap has driven interest in functional neurophysiology and network connectivity approaches that are sensitive to subtle post-injury alterations~\citep{dunkley2015a,vakorinDetectingMildTraumatic2016,antonakakisAlteredRichClubFrequencyDependent2017,vergaraDetectionMildTraumatic2017,wangDisruptedGammaSynchrony2017,allenMagnetoencephalographyAbnormalitiesAdult2021}.
Magnetoencephalography (MEG) and electroencephalography (EEG) offer millisecond-level insights into brain activity. These techniques have uncovered consistent patterns of abnormality in mTBI patients. Research shows increased slow-wave activity in these patients compared to healthy controls. Changes in resting-state network connectivity have also been documented~\citep{huangSinglesubjectbasedWholebrainMEG2014,vakorinDetectingMildTraumatic2016,vergaraDetectionMildTraumatic2017}.
These researches motivate quantitative biomarkers that can generalise across subjects and sites (or systems) while remaining transparent and clinically interpretable~\citep{grossGoodPracticeConducting2013,nisoBrainstormPipelineAnalysis2019,boonLongitudinalConsistencySourcespace2021,ferranteFLUXPipelineMEG2022}.

A substantial literature applies machine learning (ML) to neurophysiology and multimodal data for mTBI detection~\citep{vakorinDetectingMildTraumatic2016,italinnaUsingNormativeModeling2023,vergaraDetectionMildTraumatic2017,mcnerneyObjectiveClassificationMTBI2019,thanjavurRecurrentNeuralNetworkbased2021,vivaldiEvaluatingPerformanceEEG2021,caiolaEEGClassificationTraumatic2023,wallDeepLearningbasedApproach2022}.
Examples include normative modelling and support vector machines (SVMs) on MEG source features~\citep{italinnaUsingNormativeModeling2023}, recurrent neural networks on raw resting-state EEG~\citep{thanjavurRecurrentNeuralNetworkbased2021}, deep audio classifiers for voice-based screening~\citep{wallDeepLearningbasedApproach2022}, and combined frontopolar EEG with self-reported symptoms~\citep{mcnerneyObjectiveClassificationMTBI2019}. Beyond electrophysiology, resting-state functional network connectivity and diffusion MRI (fractional anisotropy) have achieved encouraging performance with classical ML~\citep{vergaraDetectionMildTraumatic2017}. Reported accuracies in these studies typically range between about 0.79 and 0.93~\citep{italinnaUsingNormativeModeling2023,vergaraDetectionMildTraumatic2017,mcnerneyObjectiveClassificationMTBI2019,thanjavurRecurrentNeuralNetworkbased2021,wallDeepLearningbasedApproach2022}, also see~\Cref{tab:mTBI_accuracy_comparison}. However, these advances in black-box AI approaches encounter significant obstacles when it comes to clinical implementation and medico-legal acceptance~\citep{rudinStopExplainingBlack2019,leslieExplainingDecisionsMade2020,WHOAIHealth2021}.
In particular, under current UK practice, automated, non-transparent models are unlikely to be accepted as standalone evidence in court; expert opinion requires methods that are explainable, auditable, and grounded in established statistical principles~\citep{lawcommissionExpertEvidence2011}.
At the same time, purely statistical models that focus on interpretability often sacrifice predictive accuracy, say 83\% in~\citet{huangSinglesubjectbasedWholebrainMEG2014} and 79\% in~\citet{italinnaUsingNormativeModeling2023}. This challenge highlights the importance of developing statistical methods that balance interpretability with strong predictive performance.~\citep{breimanStatisticalModelingTwo2001,barredoarrietaExplainableArtificialIntelligence2020,bollaertsQuantileRegressionMonotonicity2006}.

We bridge this gap by proposing an interpretable framework based on Wasserstein–Fréchet regression~\citep{petersenWasserstein$F$testsConfidence2021} of quantile functions derived from source-space power of MEG signals.
Rather than summarising the epoch magnitudes with low-dimensional moments~\citep{liBayesianInferenceGeneral2025}, we treat the entire empirical distribution of source magnitudes within a region and frequency band as the response variable.
Quantile functions preserve full distributional information and admit a natural geometry via the Wasserstein metric; conditional Fréchet means under this metric can be expressed as weighted averages of quantile functions, yielding estimators that are easy to compute and explain~\citep{petersenWasserstein$F$testsConfidence2021,petersenFunctionalDataAnalysis2016,petersenModelingProbabilityDensity2022,ramsayFunctionalDataAnalysis2006,wangFunctionalDataAnalysis2016}.
In our pipeline, epoch-wise \( L_{1} \)-constrained source magnitude imaging (VESTAL) produces robust regional magnitude distributions~\citep{huangSinglesubjectbasedWholebrainMEG2014}. Diffusion-based kernel density estimation~\citep{pelzDiffusionbasedKernelDensity2023} with fixed-point bandwidth selection provides stable density and quantile estimates. Global Wasserstein regression then links these distributional responses to covariates (age, gender), producing group- and covariate-conditioned “prototype” distributions. For an individual, classification reduces to comparing Wasserstein distances between the subject’s quantile function and covariate-matched group prototypes, yielding a transparent decision rule across quantiles and regions.

This design has several implications for accuracy and translation. First, modelling full distributions mitigates information loss from ad hoc summary statistics and captures heteroscedastic and skewed changes commonly observed post-injury~\citep{huangSinglesubjectbasedWholebrainMEG2014,allenMagnetoencephalographyAbnormalitiesAdult2021,vakorinDetectingMildTraumatic2016}. Moreover, our nonparametric, distribution-free density estimation approach makes no assumptions on the distribution of the data, allowing the method to adapt flexibly to arbitrary distribution that may arise in neuroimaging applications.
Second, the Wasserstein geometry inherently captures mass transport between distributions, which improves sensitivity to changes in slow and gamma frequency bands previously linked to mTBI~\citep{huangSinglesubjectbasedWholebrainMEG2014,vakorinDetectingMildTraumatic2016}.
Third, the decision process remains fully transparent: practitioners can examine distances to covariate-matched group means and quantile-level contributions, providing clear advantages for clinical reporting and medico-legal review.

The remainder of the paper is organised as follows. Section~\ref{sec:Methodology} formulates the Wasserstein–Fréchet regression on quantile functions and details density/quantile estimation via diffusion-based kernel density estimator (KDE). Section~\ref{sec:simulation_study} presents simulations comparing the proposed estimator to alternatives and evaluates classification performance relative to standard linear baselines. Section~\ref{sec:real_data_analysis} applies the full MRI+MEG pipeline to Cam-CAN controls~\citep{cam-canCambridgeCentreAgeing2014} and a  mTBI cohort, including multiband analyses (delta, slow gamma, fast gamma) and cross-validated evaluation. Section~\ref{sec:conclusion} discusses limitations, generalisability, and the translational potential of interpretable, distributional MEG biomarkers for mTBI.

\section{Methodology}\label{sec:Methodology}

In many applications, probability density functions arise as functional response objects in regression with Euclidean covariates.~\citet{petersenWasserstein$F$testsConfidence2021} develop a regression framework for density-valued responses under the Wasserstein metric with vector predictors. Their global model requires no local smoothing or other tuning parameters. We summarise the elements relevant to our setting and then detail the Wasserstein–Fréchet regression used here.

Let \(f_i\) denote the probability density function for subject \(i=1,\dots,n\). Each subject \(i\) provides a sample \(\mathbf{x}_i=(x_{i1},\dots,x_{i n_i})^\top\) drawn independently from \(f_i\). Sample sizes \(n_i\) may vary between subjects. We also record covariates \(\mathbf{z}_i=(z_{i1},\dots,z_{ip})^\top\) such as age and gender. Each subject has a binary clinical status label \(y_i\in\{0,1\}\). The complete training data consists of \(\{(y_i,\mathbf{x}_i,\mathbf{z}_i)\}_{i=1}^n\). For a new subject, we observe \((\mathbf{x}_0,\mathbf{z}_0)\) and aim to predict the corresponding \(y_0\).

Let \(\mathcal{D}\) denote the set of univariate probability density functions \(f\) on \(\mathbb{R}\) that are absolutely continuous and have finite second moment, i.e., \(\int_{\mathbb{R}} u^{2} f(u)\,\mathrm{d}u < \infty\). For \(f,g \in \mathcal{D}\), let \(\mathcal{M}_{f,g}\) be the set of measurable maps \(M:\mathbb{R}\to\mathbb{R}\) such that, if \(U \sim f\), then \(M(U) \sim g\). The squared Wasserstein distance between \(f\) and \(g\) is defined by

\begin{equation*}
    d_{\mathrm{W}}^{2}(f, g) \coloneqq \inf_{M \in \mathcal{M}_{f,g}} \int_{\mathbb{R}} \left(M(u) - u\right)^2 f(u) \mathrm{d} u.
\end{equation*}

In one dimension, we know the infimum has a specific solution. The optimal transport map is \(M_{f,g}^{\textrm{opt}} = G^{-1} \circ F\). Here, \(F\) and \(G\) represent the cumulative distribution functions (CDFs) of \(f\) and \(g\). This leads to a closed-form expression:
\begin{equation}\label{eq: wass2}
    d_{\mathrm{W}}^{2}(f, g) = \int_\mathbb{R} \left(M_{f,g}^{\textrm{opt}}(u) - u\right)^2 f(u) \mathrm{d} u  = \int_0^1 \left(F^{-1}(t) - G^{-1}(t)\right)^2\mathrm{d} t,
\end{equation}
where the second equality follows from the change of variables \(t = F(u)\)Here, \(F^{-1}\) and \(G^{-1}\) are the quantile functions for the two distributions.
This shows that in one dimension, the Wasserstein distance becomes the
\(L^{2}\)-distance between quantile functions.

\subsection{Random densities}\label{random-densities}

A random density \(\mathfrak{F}\) is defined as a random variable that takes values almost surely in \(\mathcal{D}\)~\citep{petersenWasserstein$F$testsConfidence2021}. The associated random CDF  and quantile function are denoted as \(F\) and \(Q = F^{-1}\) respectively~\citep{parzenNonparametricStatisticalData1979}. For clarity, we use specific notation conventions throughout this work. We use \(u,v \in \mathbb{R}\) as arguments for densities and CDFs. In contrast, we use \(s,t \in [0,1]\) as arguments for quantile and quantile-density functions. The Wasserstein-Fréchet mean and variance of \(\mathfrak{F}\) are defined as follows:
\begin{equation}\label{eq: wMeanVar}
    f_{\oplus}^{*} \coloneqq \argmin_{f \in \mathcal{D}} E\left(d_{\mathrm{W}}^{2}(\mathfrak{F}, f)\right), \quad \mathrm{Var}_\oplus(\mathfrak{F})\coloneqq E\left(d_{\mathrm{W}}^{2}(\mathfrak{F}, f_{\oplus}^{*})\right).
\end{equation}

In regression applications, we often need to model how \(\mathfrak{F}\) changes with predictor variables \(Z \in \mathbb{R}^p\). The pair \((Z, \mathfrak{F})\) follows a joint distribution \(\mathcal{G}\) on \(\mathbb{R}^p \times \mathcal{D}\). The quantities in~\eqref{eq: wMeanVar} represent the overall Fréchet mean and variance of \(\mathfrak{F}\) without conditioning on \(Z\). Let \(\mathfrak{S}_Z\) be the support of \(Z\)'s marginal distribution. We want to find the Fréchet regression function. This maps each covariate value to its corresponding conditional Fréchet mean:
\begin{equation}\label{eq:condWMean}
    f_{\oplus}(\mathbf{z}) := \argmin_{f \in \mathcal{D}} E\left[d_{\mathrm{W}}^{2}(\mathfrak{F}, f) | Z = \mathbf{z}\right], \quad \mathbf{z} \in \mathfrak{S}_Z.
\end{equation}

Next, we define the conditional Fréchet variance. Let \(F_{\oplus}(\mathbf{z})\) be the CDF that corresponds to \(f_{\oplus}(\mathbf{z})\). We denote its quantile function as \(Q_{\oplus}(\mathbf{z})\). To clarify our notation: \(f_{\oplus}(\mathbf{z},u)\) represents the value of the conditional mean density \(f_{\oplus}(\mathbf{z})\) evaluated at point \(u \in \mathbb{R}\). Similarly, \(F_{\oplus}(\mathbf{z},u)\) and \(Q_{\oplus}(\mathbf{z},t)\) denote the corresponding CDF and quantile function values at \(u\) and \(t \in [0,1]\), respectively.

For a pair \((Z,\mathfrak{F})\), we define an optimal transport map \(T_{\mathbf{z}}:\mathbb{R}\to\mathbb{R}\) for each \(\mathbf{z}\). This map transports from \(f_{\oplus}(\mathbf{z})\) to the random density \(\mathfrak{F}\). We express it as
\(T_{\mathbf{z}}(u) \coloneqq Q\!\left(F_{\oplus}(\mathbf{z},u)\right)\),
where \(Q\) represents the random quantile function of \(\mathfrak{F}\).
From equation~\eqref{eq: wass2}, we know that \(E\!\left\{Q(t)\mid Z=\mathbf{z}\right\} = Q_{\oplus}(\mathbf{z},t)\). This relationship tells us that \(E\!\left\{T_{\mathbf{z}}(u)\mid Z=\mathbf{z}\right\} = u\) for all \(u\) where \(f_{\oplus}(\mathbf{z},u) > 0\).
We can now express the conditional Fréchet variance as
\begin{equation}
    \begin{split}
        \mathrm{Var}_{\oplus}(\mathfrak{F}|Z= \mathbf{z}) & = E\left[d_{\mathrm{W}}^{2}(\mathfrak{F}, f_{\oplus}(\mathbf{z}))|Z = \mathbf{z}\right] = \int_\mathbb{R} E\left[(T_{\mathbf{z}}(u) - u)^2|Z=\mathbf{z}\right]f_{\oplus}(\mathbf{z},u)\mathrm{d} u \\
                                                          & = \int_\mathbb{R} \mathrm{Var}(T_{\mathbf{z}}(u)|Z=\mathbf{z}) f_{\oplus}(\mathbf{z},u) \mathrm{d} u.
    \end{split}
\end{equation}

The approach above requires that both marginal and conditional Wasserstein mean densities exist and have unique solutions. This property does not hold automatically in all cases. However, the sufficient conditions (A1) through (A3) established by~\citet{petersenWasserstein$F$testsConfidence2021} ensure these requirements are met. We work under these assumptions and move directly to the regression framework.

\subsection{Global Wasserstein-Fréchet regression}\label{global-wassersteinfruxe9chet-regression}

To examine the effects of covariates \(Z\) on conditional Wasserstein means, we use a global regression approach for the conditional means \(f_{\oplus}(\mathbf{z})\) from~\eqref{eq:condWMean}. Following~\citet{petersenFrechetRegressionRandom2019}, this Fréchet regression model uses a weighted Fréchet mean structure:
\begin{equation}\label{eq:Wmodel}
    f_{\oplus}(\mathbf{z}) = \argmin_{f \in \mathcal{D}} E\!\left[s(Z, \mathbf{z})\, d_{\mathrm{W}}^{2}(\mathfrak{F}, f)\right],
\end{equation}
where \( s(Z, \mathbf{z}) = 1 + (Z - \mu)^\top \Sigma^{-1}(\mathbf{z} - \mu), \quad \mu = E(Z), \ \Sigma = \mathrm{Var}(Z)\) is the weight function, and \(\Sigma\) is positive definite. This model extends linear regression to work with density-valued responses. Instead of using \(Y\) and the usual Euclidean space \((\mathbb{R}, |\cdot|)\), we substitute \(\mathfrak{F}\) and the Wasserstein space \((\mathcal{D}, d_{\mathrm{W}})\).
Model~\eqref{eq:Wmodel} gives us a way to specify the conditional Wasserstein mean of \(\mathfrak{F}\). In what follows, we concentrate on this mean specification.

\subsection{Estimation}\label{estimation}

To estimate the regression function \(f_{\oplus}(\mathbf{z})\), we use an empirical weighted least squares criterion in Wasserstein space as in~\eqref{eq:Wmodel}. We calculate the sample mean \(\bar{Z} = n^{-1}\sum_{i=1}^{n} Z_i\) and the sample covariance matrix \(\hat{\Sigma} = n^{-1}\sum_{i=1}^{n} (Z_i - \bar{Z})(Z_i - \bar{Z})^\top\).
We then define empirical weights as \(s_{in}(\mathbf{z}) = 1 + (Z_i - \bar{Z})^\top \hat{\Sigma}^{-1}(\mathbf{z} - \bar{Z})\). Let \(\mathfrak{Q}\) denote the set of quantile functions in \(L^{2}[0,1]\). Using \({\lVert \cdot \rVert}_{L^{2}}\) as the standard Hilbert norm on \(L^{2}[0,1]\), we estimate the conditional mean quantile function \(Q_{\oplus}(\mathbf{z})\) by solving:
\begin{equation}\label{eq:Qfit}
    \hat{Q}_{\oplus}(\mathbf{z}) = \argmin_{Q \in \mathfrak{Q}} \sum_{i = 1}^{n} s_{in}(\mathbf{z}) {\left\lVert Q - Q_i \right\rVert}_{L^{2}}^{2}.
\end{equation}
where \(Q_i\) is the subject-specific quantile function.

The above introduction of Wasserstein-Fréchet regression are summarised based on~\citet{petersenWasserstein$F$testsConfidence2021}.  Implementation details are provided in Algorithm~1 of~\citet{petersenWasserstein$F$testsConfidence2021}, the \textbf{R} package \textbf{WRI} for Wasserstein-Fréchet regression is available online~\href{https://cran.r-project.org/web/packages/WRI/index.html}{here}. In finite samples, \(\hat{Q}_{\oplus}(\mathbf{z})\) need not be strictly increasing and thus may fail to induce a density. However, the Lemma~2 in~\citet{petersenWasserstein$F$testsConfidence2021} guarantees that, with high probability for large samples, \(\hat{Q}_{\oplus}(\mathbf{z})\) is strictly increasing.

Having obtained the conditional mean quantile estimator \(\hat{Q}_{\oplus}(\mathbf{z})\) from the weighted Fréchet criterion in~\eqref{eq:Qfit}, we fit two separate Wasserstein–Fréchet regression models. We train one model using healthy controls and another using mTBI patients. We restrict the optimization in~\eqref{eq:Qfit} to subjects within each group \(g \in \{\mathrm{ctl}, \mathrm{mTBI}\}\). This produces covariate-matched prototype quantiles \(\hat{Q}_{\oplus}^{g}(\mathbf{z})\) that capture the distributional characteristics at covariate value \(\mathbf{z}\).
For a new subject with covariates \(\mathbf{z}_{0}\) and estimated quantile function \(\hat{Q}(\mathbf{z}_{0})\), we classify by comparing Wasserstein distances. We compute distances between \(\hat{Q}(\mathbf{z}_{0})\) and both prototypes: \(\hat{Q}_{\oplus}^{\mathrm{ctl}}(\mathbf{z}_{0})\) and \(\hat{Q}_{\oplus}^{\mathrm{mTBI}}(\mathbf{z}_{0})\). Algorithm~\ref{alg:classification} provides the complete procedure and decision rule. We select the classification threshold through cross-validation on the training set.

\begin{algorithm}[t]
    \caption{Covariate-matched classification via Wasserstein distances between quantile functions}\label{alg:classification}
    \begin{algorithmic}[1]
        \Require Fitted group-specific conditional prototype quantiles
        \(\hat{Q}_{\oplus}^{\mathrm{ctl}}(\mathbf{z})\) and \(\hat{Q}_{\oplus}^{\mathrm{mTBI}}(\mathbf{z})\) from~\eqref{eq:Qfit};
        subject covariates \(\mathbf{z}_{0}\);
        subject data (MEG/MRI) to estimate \(\hat{Q}(\mathbf{z}_{0})\);
        threshold \(k\) selected by cross-validation (or grid search).
        \Ensure Predicted label \(\hat{y}_{0}\in\{\mathrm{control},\mathrm{mTBI}\}\); distances \(d_{1}, d_{2}\).
        \State Estimate the subject-specific quantile function \(\hat{Q}(\mathbf{z}_{0})\) using diffusion-based KDE and monotone refit (\Cref{subsec:Quantile_Function_Estimation}).
        \State Compute covariate-matched group prototypes at \(\mathbf{z}_{0}\):
        \(\hat{Q}_{\oplus}^{\mathrm{ctl}}(\mathbf{z}_{0})\) and \(\hat{Q}_{\oplus}^{\mathrm{mTBI}}(\mathbf{z}_{0})\) via~\eqref{eq:Qfit}.
        \State Compute Wasserstein distances using~\eqref{eq: wass2}:
        \[
            d_{1} \gets \Big(\!\int_{0}^{1}\!\big[\hat{Q}(\mathbf{z}_{0})-\hat{Q}_{\oplus}^{\mathrm{ctl}}(\mathbf{z}_{0})\big]^{2}\mathrm{d}t\Big)^{1/2},\quad
            d_{2} \gets \Big(\!\int_{0}^{1}\!\big[\hat{Q}(\mathbf{z}_{0})-\hat{Q}_{\oplus}^{\mathrm{mTBI}}(\mathbf{z}_{0})\big]^{2}\mathrm{d}t\Big)^{1/2}.
        \]
        \State Decision rule:
        \[
            \hat{y}_{0} \gets
            \begin{cases}
                \mathrm{control}, & \text{if } d_{1} \le k\, d_{2}, \\
                \mathrm{mTBI},    & \text{otherwise}.
            \end{cases}
        \]
        \State Select \(k\) by cross-validation to maximise balanced \( F_{1} \) score.
    \end{algorithmic}
\end{algorithm}

The theoretical framework of~\citet{petersenWasserstein$F$testsConfidence2021} assumes that subject-level densities and quantile functions are directly observable. However, real neuroimaging studies must estimate these functions from finite, noisy data samples. This creates several practical challenges. We must ensure smoothness and proper boundary behaviour. Most importantly, we need to guarantee that estimated quantile functions remain monotonic.
We tackle these problems using a two-step approach in~\Cref{subsec:Quantile_Function_Estimation}. First, we estimate densities through diffusion-based kernel density estimation with fixed-point bandwidth selection. Second, we enforce quantile monotonicity using a lightweight quadratic programming procedure. These steps produce valid CDF and quantile function pairs that work well with Wasserstein-Fréchet regression and classification methods.

\subsection{Quantile Function Estimation}\label{subsec:Quantile_Function_Estimation}
In this section, we estimate quantile functions using diffusion-based kernel density estimation~\citep{botevKernelDensityEstimation2010}. For clarity, we temporarily omit the subject index \( i \) in the observations \( \mathbf{x}_{i}= (x_{i1},x_{i2},\ldots,x_{n_{i}})^{\top} \), and let \( \mathbf{x}= (x_{1},x_{2},\ldots,x_{n})^{\top} \) when calculating the quantile estimation. The kernel density estimator (KDE) is a non-parametric method for estimating the probability density function of \( \mathbf{x} \). The KDE of the true density \( f \) is defined as
\begin{equation}\label{eq:non-parametric_kde}
    \hat{f}(x;h) = \frac{1}{n h} \sum_{i=1}^{n} K\left(\frac{x - x_{i}}{h}\right),
\end{equation}
where \( K \) is the kernel function, \( h >0\) is the bandwidth, and \( n \) is the number of observations. The kernel \( K \) is a symmetric probability density that integrates to 1; common choices of kernel \( K \) include Gaussian, Epanechnikov, and uniform kernels. However, standard KDE has several problems. First, data-driven plug-in bandwidth selectors often assume normality, which is unrealistic in practice. Second, the Gaussian kernel lacks local adaptivity~\citep{terrellVariableKernelDensity1992}. Third, most KDEs suffer boundary bias when the random variable has bounded support~\citep{marronTransformationsReduceBoundary1994,parkAdaptiveVariableLocation2003}.

The diffusion-based KDE addresses these issues. The basic strategy relies on approximating~\eqref{eq:non-parametric_kde} via linear diffusion process smoothing~\citep{botevKernelDensityEstimation2010}:
\begin{equation}\label{eq:linear_diffusion_PDE}
    \frac{\partial  }{\partial h}f(x;h) = Lf(x;h),\qquad x\in \mathcal{X},\ h>0,
\end{equation}
where \( \mathcal{X} \) is the domain of \( x \). The second-order linear differential operator \( L \) is defined as \( \frac{1}{2}\frac{\mathrm{d} }{\mathrm{d} x}(a(x)\frac{\mathrm{d} }{\mathrm{d} x}(\frac{\cdot}{p(x)})) \), where \( a(x) \) and \( p(x) \) are arbitrary positive functions on \( \mathcal{X} \).
Equation~\eqref{eq:linear_diffusion_PDE} describes an Itô diffusion in the pseudo-time \( h \), and its solution yields the evolving density of \( x \). In this paper we take \( a(x)=p(x)\propto 1 \), in which case the solution of \eqref{eq:linear_diffusion_PDE} recovers the Gaussian-kernel KDE in \eqref{eq:non-parametric_kde}. Next, we describe how to choose the bandwidth \( h \) in the diffusion-based KDE. Let \( t=h^{2} \). Using a first-order Taylor expansion, the asymptotic mean integrated squared error (AMISE) is \( t^{2}{\left\lVert f^{\prime\prime} \right\rVert}^{2}/4+1/(2n\sqrt{\pi t}) \)~\citep{botevKernelDensityEstimation2010,wandKernelSmoothing1995}. The asymptotically optimal bandwidth is
\begin{equation}\label{eq:optimal_bandwidth}
    _{*}t = \left(\frac{1}{2n\sqrt{\pi}{\left\lVert f^{\prime\prime} \right\rVert}^{2}}\right)^{2/5}.
\end{equation}
However, \( f^{\prime\prime} \) is unknown in practice. Within diffusion-based KDE, the \( L^{2} \) norm of the \( j \)-th derivative of \( f \) can be expressed as
\begin{equation}\label{eq:diffusion_kde}
    {\left\lVert f^{(j)} \right\rVert}^{2} = \frac{(-1)^{j}}{n^{2}}\sum_{k=1}^{n}\sum_{m=1}^{n}\phi^{(2j)}(x_{k},x_{m};2t_{j}),
\end{equation}
where \( \phi^{(2j)}(x_{k},x_{m};2t_{j}) \) is the \( 2j \)-th derivative of the Gaussian kernel \( K(x_{k},x_{m};2t_{j}) \) with respect to \( x_{k} \) and \( x_{m} \), and \( t_{j} \) is the bandwidth for the \( j \)-th derivative~\citep[][eq. (26)]{botevKernelDensityEstimation2010}. Computing~\eqref{eq:diffusion_kde} requires \( O(n^{2}) \) operations, making it too slow for large \( n \). To reduce complexity,~\citet{botevKernelDensityEstimation2010} exploit the discrete Fourier transform (DFT) to compute the \( L^{2} \)-norm of the \( j \)-th derivative. For the Gaussian kernel, the Fourier transform is also Gaussian, i.e., \( \mathcal{F}[\phi](\omega,t) = e^{-\omega^{2}t/2}.\)
The Fourier transform of \( \hat{f}(x,t) \) is therefore \(\mathcal{F}[\hat{f}](\omega,t)=\mathcal{F}[\phi](\omega,t)\cdot  \hat{f}_{\text{emp}}(\omega)\)
where \( \hat{f}_{\text{emp}}(\omega) =\frac{1}{n}\sum_{i=1}^{n}e^{-i\omega x_{i}}\) is the empirical characteristic function. The power spectrum is
\({\left\lvert \mathcal{F}[\hat{f}](\omega,t) \right\rvert}^{2} = e^{-\omega^{2}t}{\left\lvert \hat{f}_{\text{emp}}(\omega) \right\rvert}^{2} .
\)
Parseval’s theorem states that the \( L^{2} \)-norm of a function equals the \( L^{2} \)-norm of its Fourier transform (up to a \( 2\pi \) factor). Using \( \mathcal{F}[f^{(j)}](\omega,t)=\omega^{j}\mathcal{F}[\hat{f}](\omega,t) \), we obtain
\begin{equation}\label{eq:power_spectrum_2}
    {\left\lVert f^{(j)} \right\rVert}^{2} = \frac{1}{2\pi}\int_{-\infty}^{\infty} {\left\lvert \mathcal{F}[\hat{f}^{(j)}](\omega,t) \right\rvert}^{2} \mathrm{d} \omega = \frac{1}{2\pi}\int_{-\infty}^{\infty} \omega^{2j}e^{-\omega^{2}t}{\left\lvert \hat{f}_{\text{emp}}(\omega) \right\rvert}^{2} \mathrm{d} \omega.
\end{equation}
In practice, the integral in~\eqref{eq:power_spectrum_2} is approximated using a discrete cosine transform (DCT):
\begin{itemize}
    \item Let \( v_{k} \) be the Type-II DCT coefficients of the binned data;
    \item The squared DCT coefficients \( a_{k}\coloneqq (v_{k}/2)^{2} \) approximate \( {\left\lvert \hat{f}_{\text{emp}}(\omega) \right\rvert}^{2} \) at discrete frequencies \( \omega_{k} = \frac{k\pi}{R}, k=1,2,\ldots \), where \( R \) is the data range;
    \item Let \( I_{k}=k^{2} \) denote squared frequency terms (Laplacian eigenvalues in the DCT basis);
    \item Gaussian smoothing (i.e., \( e^{-\omega^{2}t} \)) in frequency space is approximated by \( e^{-I_{k}\pi^{2}t} \).
\end{itemize}
Then, the \( L^{2} \)-norm of the \( j \)-th derivative of \( f \) with bandwidth \( t_{j} \) can be approximated as
\begin{equation}\label{eq:L2-DCT-approximate}
    {\left\lVert f^{(j)} \right\rVert}^{2} \approx 2\pi^{2j}\sum_{k} I_{k}^{j}*a_{k}*e^{-I_{k}\pi^{2}t_{j}}.
\end{equation}

In addition, Proposition 2 in~\citet{botevKernelDensityEstimation2010} shows that the optimal bandwidth \( _{*}t_{j} \) for the \( j \)-th derivative is
\begin{equation}\label{eq:optimal_bandwidth_j}
    _{*}t_{j} = \left(\frac{1+1/2^{j+1/2}}{3}\frac{1\times 3\times 5\times \cdots \times(2j-1)}{n\sqrt{\pi/2}{\left\lVert f^{(j+1)} \right\rVert}^{2}}\right)^{2/(3+2j)}.
\end{equation}
\Cref{eq:optimal_bandwidth,eq:L2-DCT-approximate,eq:optimal_bandwidth_j} motivate an \( l \)-stage algorithm to find the optimal bandwidth; see Algorithm~\ref{alg:bandwidth}.
\begin{algorithm}
    \caption{Fixed Point Computation for Searching Optimal Bandwidth}\label{alg:bandwidth}
    \begin{algorithmic}[1]
        \Require \( t \), \( n \) (number of data points), \( I_k \) (vector of squared indices of DCT), \( a_k\)  (transformed coefficients) \( k=1,2,\ldots, \)
        \Ensure \( _{*} \)t (fixed point result)

        \State Set \( l = 7 \) (initial order)
        \State Compute \( f = 2 \pi^{2l} \sum_k I_k^l \cdot a_k \cdot \exp(-I_k \pi^2 t) \)
        \For{\( s = l-1 \) to \( 2 \) (decreasing)}
        \State Compute \( m_0 =  (\prod_{j=1}^{s}2j-1)/\sqrt{2\pi} \)
        \State Compute \( c = (1 + (1/2)^{s+1/2})/3\)
        \State Compute \( t_s = \left( \frac{2 \cdot c \cdot m_0}{n \cdot f} \right)^{\frac{2}{3 + 2s}} \)
        \State Update \( f = 2 \pi^{2s} \sum_k I_k^s \cdot a_k \cdot \exp(-I_k \pi^2 \cdot t_s) \)
        \EndFor
        \State Compute the zero point \( _{*}t \) of the equation \(  t - \left( 2 n \sqrt{\pi} f \right)^{-2/5}=0 \).
    \end{algorithmic}
\end{algorithm}
We set the initial order \( l = 7 \) following the recommendation in~\citet{botevKernelDensityEstimation2010}.

Once the optimal bandwidth \( _{*}t \) is obtained, the density \( f \) is computed by smoothing \( v_{k} \) in the frequency domain. Let \( v_{k}^{*}\coloneqq v_{k}*\exp(-_{*}t(\pi k)^{2}/2) \). To construct the smoothed density functions, we apply the inverse DCT to the coefficients \( v_{k}^{*}, k=1,2,\ldots \). Furthermore, we need to normalise the smoothed density to integrate to 1. To estimate the cumulative distribution function (CDF), we reuse the DCT representation. With the assumption of a Gaussian kernel, the optimal bandwidth for smoothing the distribution function~\citep{bowmanBandwidthSelectionSmoothing1998} is
\begin{equation}\label{eq:optimal_bandwidth_cdf}
    _{*}t_{cdf} = \left(\frac{1}{\sqrt{\pi} n \left\Vert f^{\prime} \right\Vert^{2}}\right)^{2/3}.
\end{equation}
By substituting \( j=1 \) and \( t_j =  \,_*t \) into~\eqref{eq:L2-DCT-approximate}, we directly obtain \( \left\Vert f^{\prime} \right\Vert^{2} \). Hence, \( _{*}t_{cdf} \) can be computed via~\eqref{eq:optimal_bandwidth_cdf}. The CDF estimation procedure is given in Algorithm~\ref{alg:cdf_estimation}.
\begin{algorithm}
    \caption{CDF Estimation via Diffusion-based KDE}\label{alg:cdf_estimation}
    \begin{algorithmic}[1]
        \State Compute $f = 2\pi^2 \sum_{k} I_k \cdot a_k \cdot \exp(-_{*}tI_k \pi^2)$
        \State Calculate $t_{\text{cdf}} = (\sqrt{\pi} \cdot f \cdot N)^{-2/3}$
        \State Smooth the discrete cosine transform coefficients:
        \Statex $a_{k}^{*} = a_k \cdot \exp(-k^2 \pi^2  t_{\text{cdf}}/2), k=0,1,\ldots,n-1$
        \State Apply inverse DCT and calculate cumulative sum:
        \Statex $\mathbf{p}=( p_{0},p_{1},\ldots,p_{n-1} )= \text{cumsum}(\text{idct}(a_{0}^{*},\ldots,a_{n-1}^{*}))/(n-1)$
        \State Normalise CDF to ensure it ranges from 0 to 1:
        \Statex $\text{cdf} = \mathbf{p}/p_{n-1}$
    \end{algorithmic}
\end{algorithm}
Finally, we compute the quantile function \( Q(u) = F^{-1}(u) \) by using one-dimensional linear interpolation (e.g., the function “interp1d” in the Python package “scipy”) based on binned \( \mathbf{x} \) and the estimated CDF in~\Cref{alg:cdf_estimation}. To enforce quantile monotonicity, we directly refit the quantiles via the following optimization problem:
\begin{equation*}
    \min_{\mathbf{q}^{*}} \sum_{j=1}^m \left[ t_j \left( q_j - q^{*}_j \right)\right]^2 \quad ~\text{subject to} \quad  q^{*}_{j+1} - q^{*}_{j}\geq 0, \quad j=1,\ldots,m-1,
\end{equation*}
where \( \mathbf{q}^{*}= (q_{1}^{*},q_{2}^{*},\ldots,q_{m}^{*}) \), \( t_j \) is the \( j \)-th quantile weight, \( q^{*}_j \) is the \( j \)-th refitted quantile of the estimated distribution function, and \( q_{j} \) is the \( j \)-th interpolated quantile from~\Cref{alg:cdf_estimation}. This problem can be solved using quadratic programming or Lagrange multipliers. Such monotone refitting is widely used to ensure valid quantile functions~\citep{takeuchiNonparametricQuantileEstimation2006,bollaertsQuantileRegressionMonotonicity2006,petersenWasserstein$F$testsConfidence2021}. The code of~\Cref{alg:classification,alg:bandwidth,alg:cdf_estimation} is available in the Python package~\href{https://github.com/innovision-ip/NeuroWAR}{NeuroWAR}.

\section{Simulation Study}\label{sec:simulation_study}
\subsection{Kernel Density Estimation}\label{subsec:kernel_density_estimation}

In this section, we conduct a simulation study to evaluate the performance of the proposed method. First, we benchmark the implementation against a known probability density in the context of kernel density estimation (KDE).~\citet{pelzDiffusionbasedKernelDensity2023} proposed another diffusion-based KDE (diffKDE). The primary difference concerns the selection of the optimal bandwidth: diffKDE uses a numerical approximation, whereas our method employs the fixed-point algorithm of~\citet{botevKernelDensityEstimation2010}.

Here, we use the mixture of Gaussian distributions~\citep[][eq. (39)]{pelzDiffusionbasedKernelDensity2023} as the underlying density function:
\begin{equation*}
    f(x) = 0.3\frac{1}{\sqrt{2\pi}}e^{-\frac{1}{2}(x-6)^2}+0.6\frac{1}{0.7\sqrt{2\pi}}e^{-\frac{1}{2}\left(\frac{x-9.5}{0.7}\right)^2} +0.1\frac{1}{0.5\sqrt{2\pi}}e^{-\frac{1}{2}\left(\frac{x-12}{0.5}\right)^2}.
\end{equation*}
This density is a three-component Gaussian mixture with means \( (6,9.5,12) \) and standard deviations \( (1,0.7,0.5) \). The component weights are 0.3, 0.6, and 0.1, respectively. While this simulation employs a Gaussian mixture model for illustration, we emphasize that the subsequent Wasserstein-Fréchet regression framework makes no distributional assumptions. The method is entirely distribution-free and does not require knowledge of the underlying data. We use \( g_0 \) to denote the density estimate obtained by the diffusion-based KDE of~\citet{pelzDiffusionbasedKernelDensity2023}, and \( g_1 \) to denote the density estimate from our procedure. We compare performance using the total variation distance. Let \( p \) and \( q \) be any density functions; the total variation distance is defined as
\(\delta(p,q) = \frac{1}{2} \int_{\mathbb{R}} \left| p(x) - q(x) \right| \mathrm{d} x\).
Our algorithm is implemented in the Python package~\href{https://github.com/innovision-ip/NeuroWAR}{NeuroWAR}, while the diffKDE implementation is available in the Python package~\href{https://zenodo.org/records/13736609}{diffKDE}.

\begin{figure}[ht]
    \centering
    \begin{subfigure}[bt]{0.45\textwidth}
        \centering
        \includegraphics[width=\textwidth]{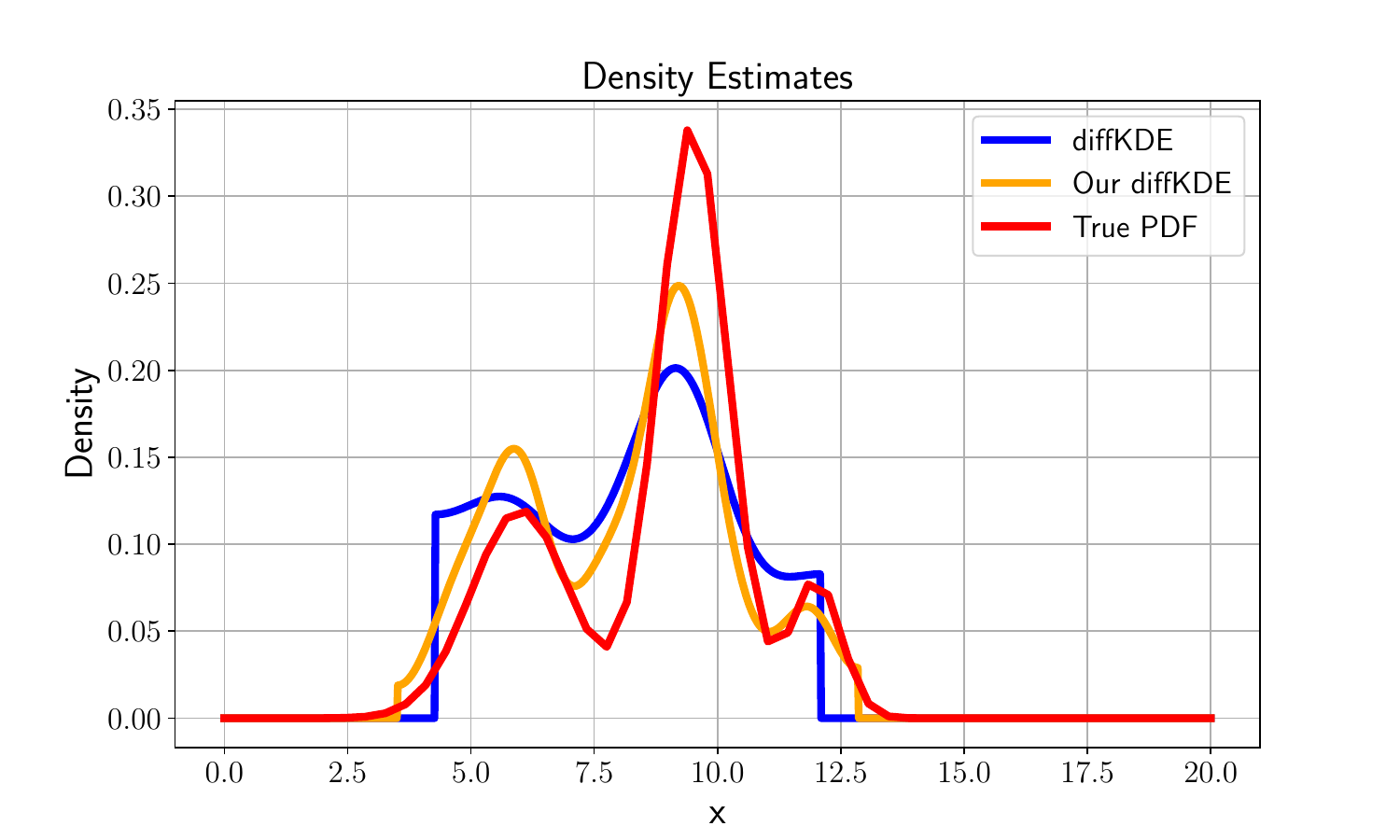}
        \caption{\( \delta(f,g_0)=0.225 \), \( \delta(f,g_1)=0.168 \)}\label{fig:figs/kde_compare_n50.pdf}
    \end{subfigure}
    \begin{subfigure}[bt]{0.45\textwidth}
        \centering
        \includegraphics[width=\textwidth]{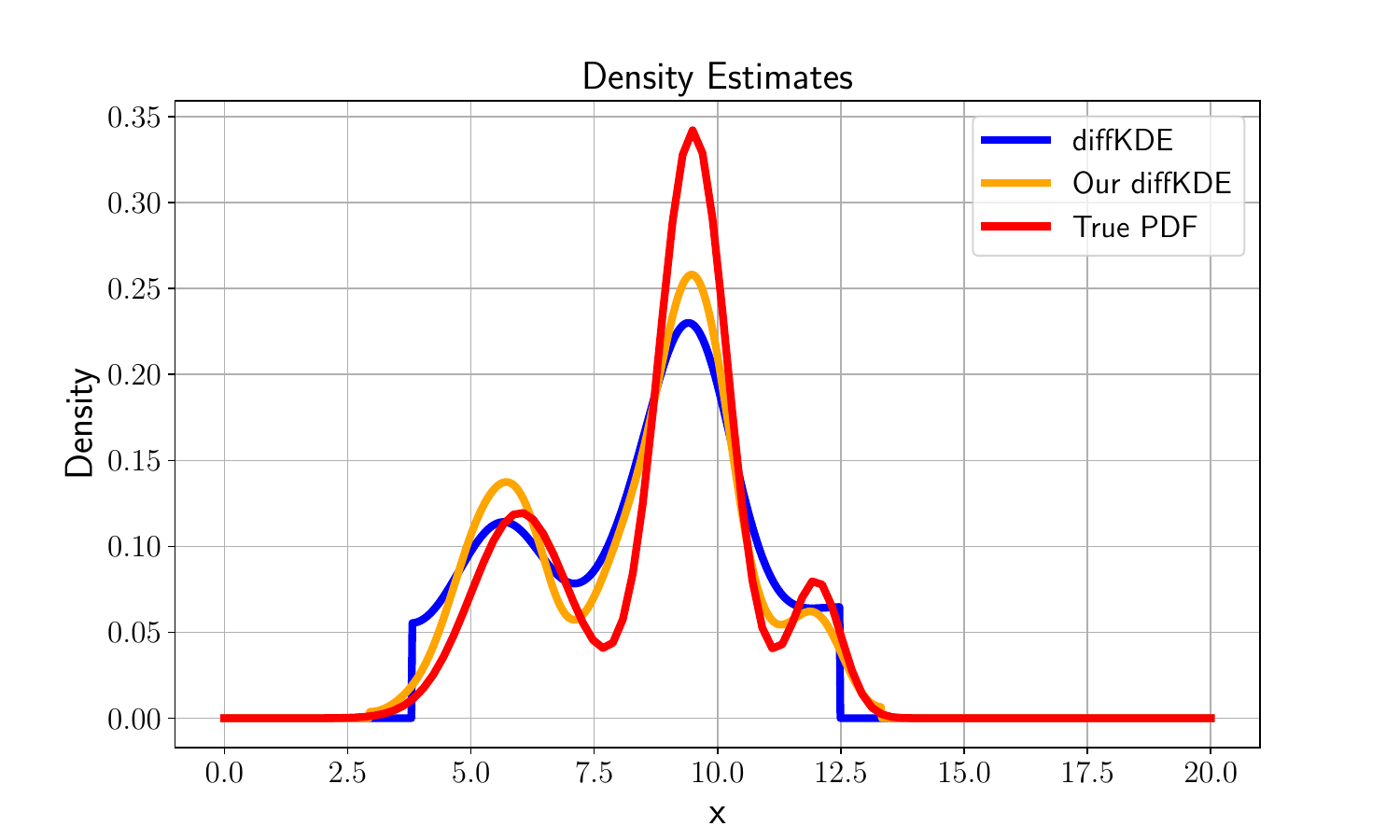}
        \caption{\( \delta(f,g_0)=0.157 \), \( \delta(f,g_1)=0.122 \)}\label{fig:figs/kde_compare_n100.pdf}
    \end{subfigure}
    \begin{subfigure}[bt]{0.45\textwidth}
        \centering
        \includegraphics[width=\textwidth]{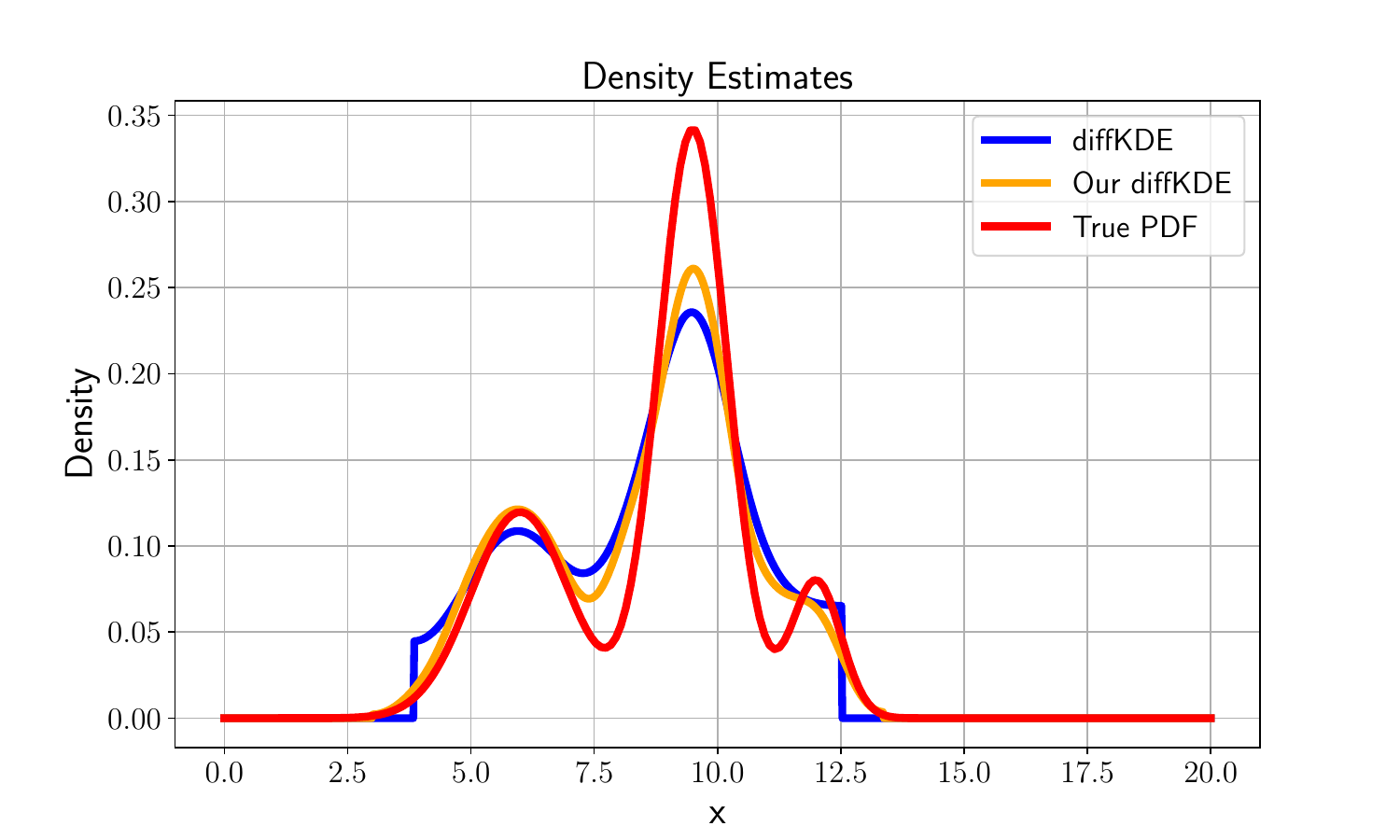}
        \caption{\( \delta(f,g_0)=0.143 \), \( \delta(f,g_1)=0.105 \)}\label{fig:figs/kde_compare_n200.pdf}
    \end{subfigure}
    \begin{subfigure}[bt]{0.45\textwidth}
        \centering
        \includegraphics[width=\textwidth]{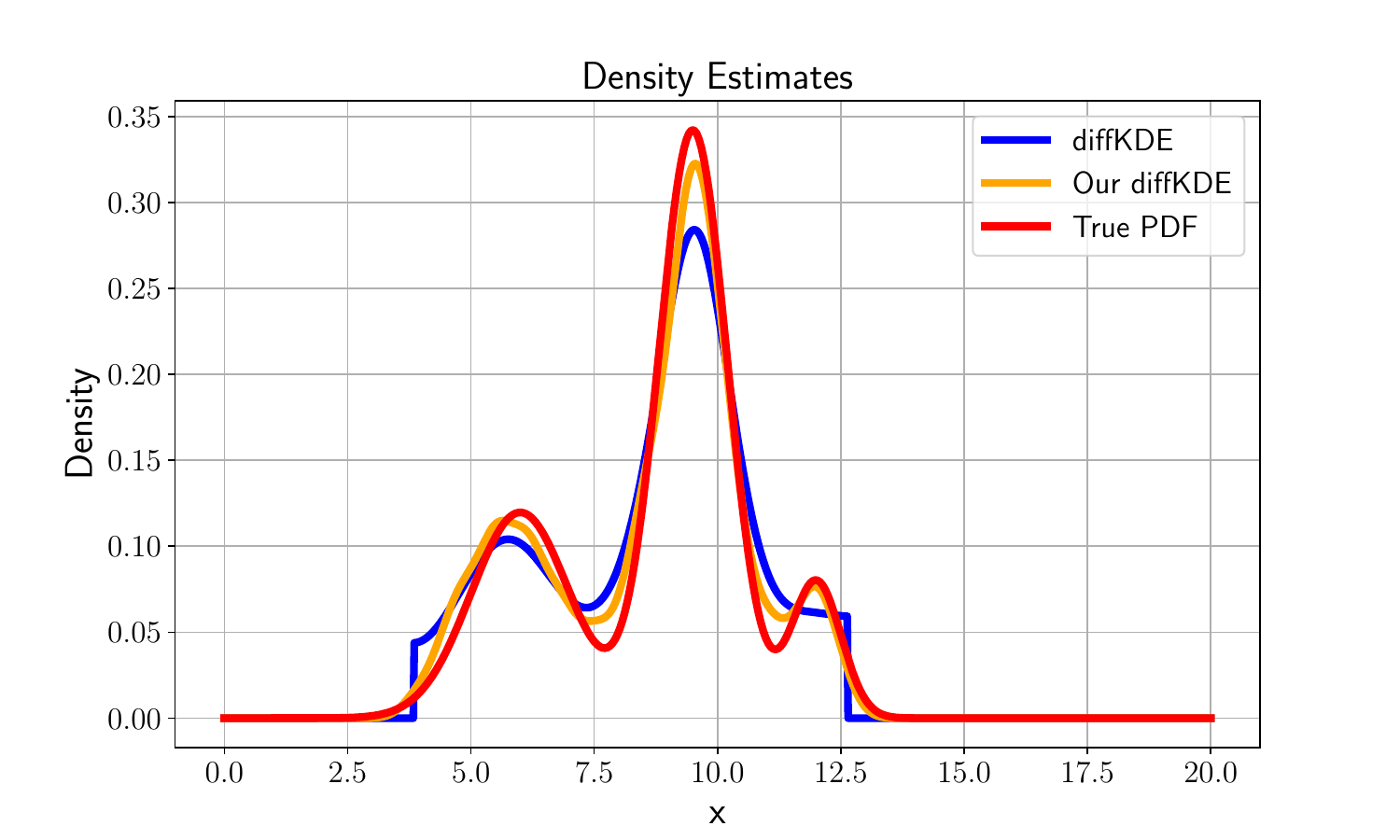}
        \caption{\( \delta(f,g_0)=0.104 \), \( \delta(f,g_1)=0.060 \)}\label{fig:figs/kde_compare_n400.pdf}
    \end{subfigure}
    \caption{Comparison of total variance distances for different kernel density estimation methods.}\label{fig:comparison_of_total_variance_distances}
\end{figure}

We consider sample sizes \( n = 50,100,200,400 \). The densities \( f \), \( g_0 \), \( g_1 \) and the corresponding distances \( \delta(f,g_0) \), \( \delta(f,g_1) \) are shown in~\Cref{fig:comparison_of_total_variance_distances}. The results indicate that our method achieves uniformly smaller total variation distance than diffKDE across all sample sizes. As expected, the distance decreases with larger \( n \), reflecting improved recovery of the underlying distribution.

\subsection{Wasserstein-Fréchet Regression}\label{subsec:WassersteinFréchet_regression}
This section presents simulation experiments to test our Wasserstein-Fréchet regression approach. We examine how well it performs compared to standard linear regression methods. First, we generate data according to the following model:
\begin{equation}\label{eq:simulation_model}
    \mu =  \beta_0 + \beta_1 \cdot \text{age}+ \beta_2 \cdot \text{gender}
\end{equation}
where \(\beta_0 \sim N(0, 0.5)\), \(\beta_1 \sim N(\nu_1,\sigma_1)\), and \(\beta_2 \sim N(\nu_2, \sigma_2)\) are mutually independent coefficients; \(\text{`age'}\) denotes the subject’s age and \(\text{`gender'}\) denotes the sex indicator (1 for male, 0 for female).~\( \nu_1, \nu_2,\sigma_1, \sigma_2 \) are the means and standard deviations of the normal distributions for \(\beta_1\) and \(\beta_2\).~\Cref{eq:simulation_model} specifies a linear relation between the response \(\mu\) and covariates \(\text{`age'}\) and \(\text{`gender'}\). Given `age' and `gender', we randomly sample the tuple \((\beta_0^{(1)},\beta_1^{(1)}, \beta_2^{(1)})\) as subject-specific coefficients. The conditional response is \( \beta_0^{(1)}+\beta_1^{(1)}\cdot \text{age} + \beta_2^{(1)} \cdot \text{gender} \), with variance \(0.25+ \sigma_1^2 \cdot \text{age}^2 + \sigma_2^2 \cdot~\text{gender}^2 \). For brevity, write \( \mu_{1|\cdot} =\beta_0^{(1)}+\beta_1^{(1)}\cdot \text{age} + \beta_2^{(1)} \cdot \text{gender}\) and \( \sigma_{1|\cdot}^{2} =0.25+ \sigma_1^2 \cdot \text{age}^2 + \sigma_2^2 \cdot~\text{gender}^2\) for the conditional mean and variance given \( \text{`age'} \) and \( \text{`gender'} \). Moreover, \( E[\mu]= \mu^{*} =\nu_1\cdot \text{age} + \nu_2 \cdot \text{gender}\), so the density \( N(\mu_{1|\cdot},\sigma_{1|\cdot}) \) can be viewed as a random mean-shift of \( N(\mu^{*},\sigma_{1|\cdot}) \). We then sample observations from \( N(\mu_{1|\cdot},\sigma_{1|\cdot}) \) conditional on each subject’s `age' and `gender'.

For the control group, we set \( \nu_1 = 0.1, \nu_2 = 2, \sigma_1 = 0.1, \sigma_2 = 0.5 \). For the mild traumatic brain injury (mTBI) group, we set \( \nu_2 = 1, \sigma_2 = 0.5 \), while \( (\nu_1, \sigma_{1}) \) vary over the Cartesian product \( \{(\nu_1, \sigma_{1}) \mid \nu_1 \in A, \sigma_{1} \in B\} \), with \( A = \{0.1, 0.3, 0.5, 0.7\} \) and \( B = \{0.1, 0.2, 0.3, 0.4, 0.5, 0.6\} \). For each group, we simulate 2000 subjects with `age' uniformly sampled from \([18, 90]\) and `gender' sampled from \(\{0,1\}\). For each subject, we randomly draw one tuple \((\beta_0^{(1)},\beta_1^{(1)}, \beta_2^{(1)})\) and then generate 1000 observations from \( N(\mu_{1|\cdot},\sigma_{1|\cdot}) \).

\begin{figure}[ht]
    \centering
    \begin{subfigure}[bt]{0.45\textwidth}
        \centering
        \includegraphics[width=\textwidth]{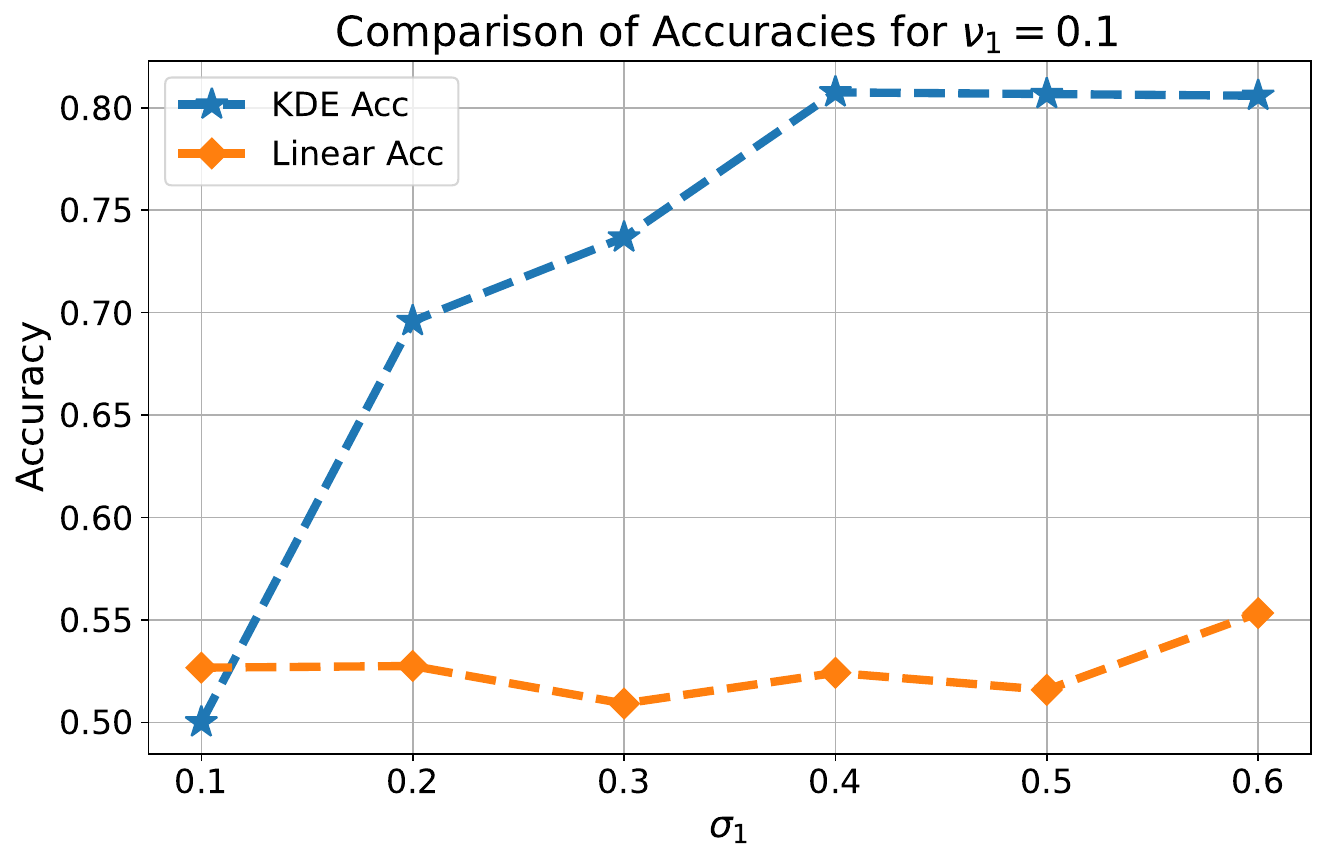}
        \caption{\( \nu_1 = 0.1 \)}\label{fig:figs/nu_1_comparison.pdf}
    \end{subfigure}
    \begin{subfigure}[bt]{0.45\textwidth}
        \centering
        \includegraphics[width=\textwidth]{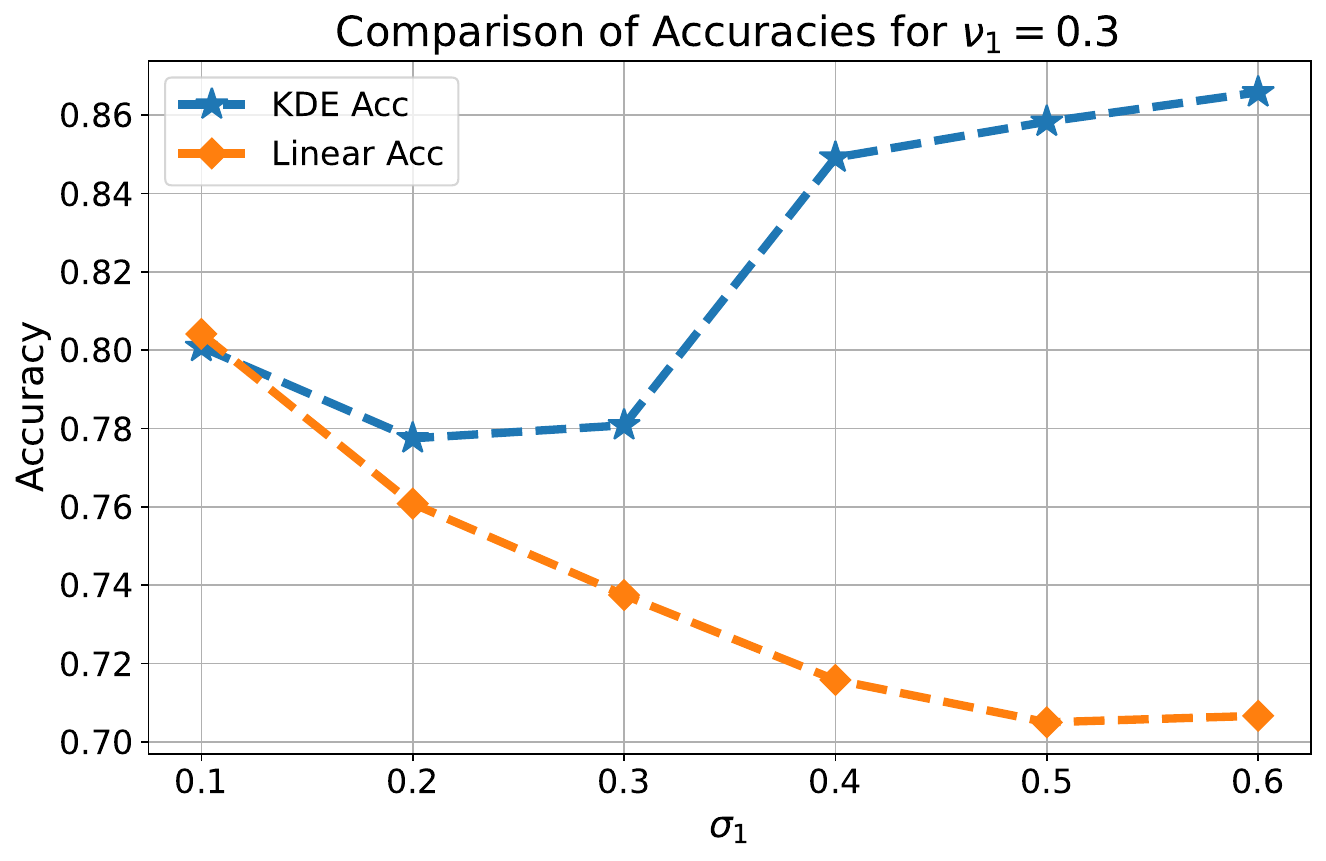}
        \caption{\( \nu_1 = 0.3 \)}\label{fig:figs/nu_2_comparison.pdf}
    \end{subfigure}
    \begin{subfigure}[bt]{0.45\textwidth}
        \centering
        \includegraphics[width=\textwidth]{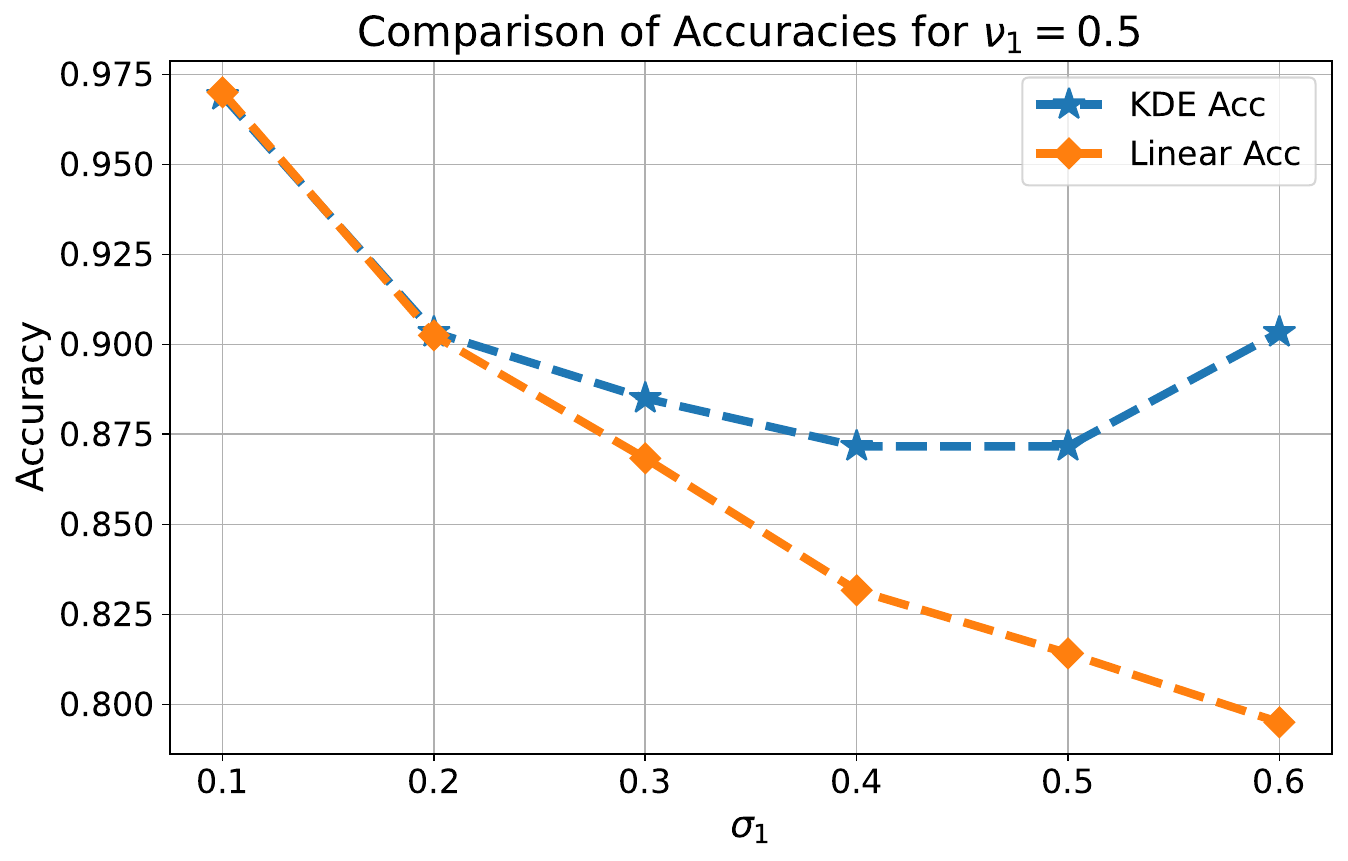}
        \caption{\( \nu_1 = 0.5 \)}\label{fig:figs/nu_3_comparison.pdf}
    \end{subfigure}
    \begin{subfigure}[bt]{0.45\textwidth}
        \centering
        \includegraphics[width=\textwidth]{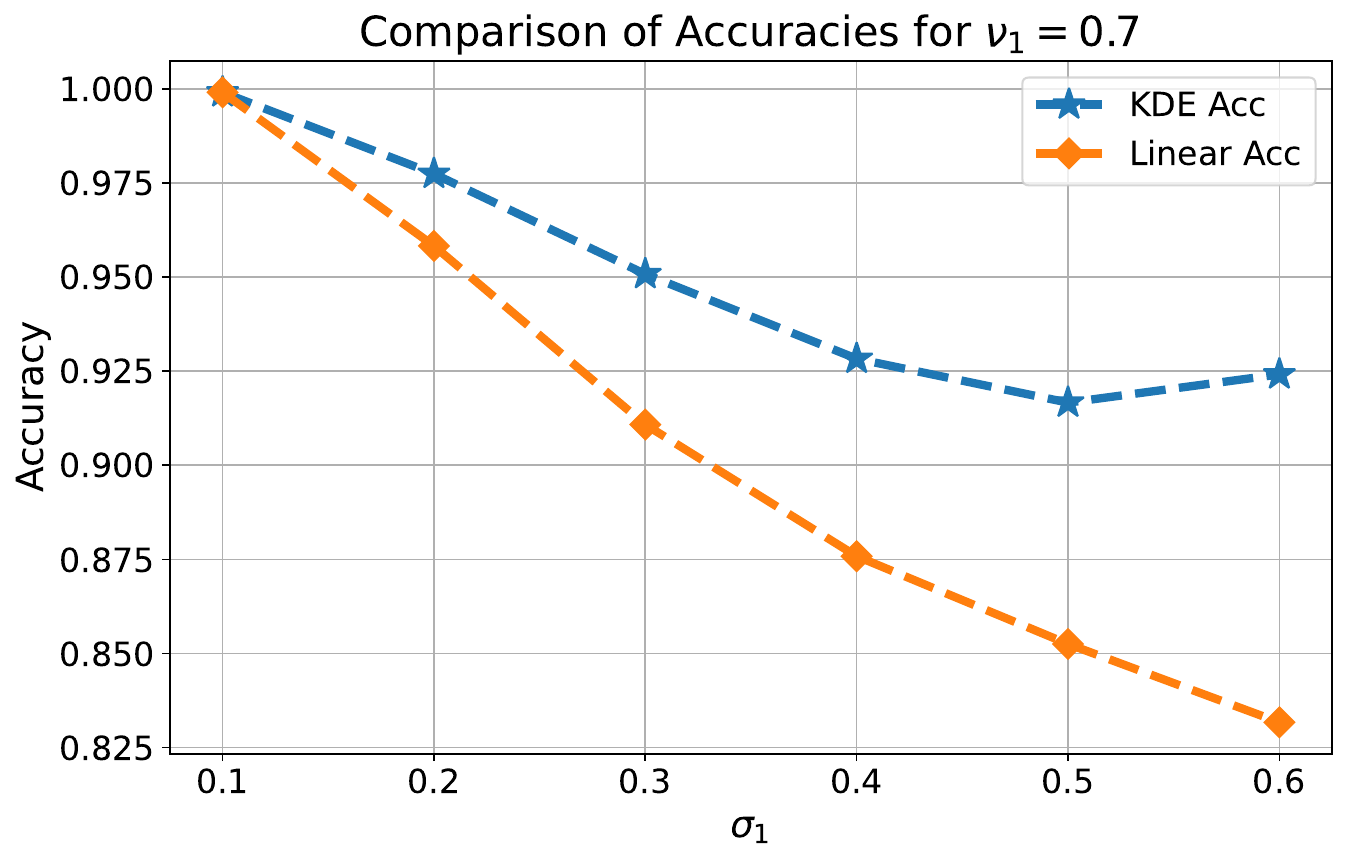}
        \caption{\( \nu_1 = 0.7 \)}\label{fig:figs/nu_4_comparison.pdf}
    \end{subfigure}
    \caption{Comparison of Accuracies for Different \( \nu_1 \) and \( \sigma_1 \) Values}\label{fig:accuracy_comparison}
\end{figure}

We apply the Wasserstein–Fréchet regression method to our simulated data following the approach outlined in Section~\ref{sec:Methodology}. To establish a performance baseline, we additionally implement a standard linear regression approach. The linear baseline operates according to this specification:

We denote \( y_{ij} \) as the observation for subject \( i \) and sample \( j \), where \( i=1,2,\ldots,m \) and \( j=1,2,\ldots,1000 \). The subject-level summary is calculated as \( \bar{y}_{i}=\sum_{j=1}^{1000}y_{ij} \), which represents the total response for each individual given their `age' and `gender'. The linear regression model is
\begin{equation}\label{eq:linear_regression_model_sample}
    \bar{y}_{i} = \alpha_0 + \alpha_1 \cdot \text{age}_i + \alpha_2 \cdot \text{gender}_i +\epsilon_i,\quad i=1,2,\ldots,m,
\end{equation}
where \( \epsilon_i \) represents zero-mean Gaussian noise with variance \( \sigma^2 \). The coefficients \( \alpha_0, \alpha_1, \alpha_2 \) are estimated by ordinary least squares (OLS). Let \( \hat{\alpha}_0^1, \hat{\alpha}_1^1, \hat{\alpha}_2^1 \) and \( \hat{\alpha}_0^2, \hat{\alpha}_1^2, \hat{\alpha}_2^2 \) be the fitted coefficients for the control and mTBI groups, respectively. For an unseen subject with `age' and `gender', the corresponding group-wise predictions are \( \hat{\bar{y}}_{1|\cdot} \) and \( \hat{\bar{y}}_{2|\cdot} \). Given the subject’s average \( \bar{y}_{0|\cdot} \), we classify this subject as control if \( {\left\lvert \hat{\bar{y}}_{1|\cdot} - \bar{y}_{0|\cdot} \right\rvert} \leq {\left\lvert \hat{\bar{y}}_{2|\cdot} - \bar{y}_{0|\cdot} \right\rvert} \), and as mTBI otherwise. We evaluate linear regression and Wasserstein-Fréchet regression using classification accuracy,  which is defined as \( ACC = (TP+TN)/(TP+TN+FP+FN) \), where \( TP \), \( TN \), \( FP \), and \( FN \) denote true positives, true negatives, false positives, and false negatives. Data splitting follows a 70/30 training-testing ratio. The implementation code of Wasserstein-Fréchet regression and linear regression classifiers is available in the Python package~\href{https://github.com/innovision-ip/NeuroWAR}{NeuroWAR}. The accuracies for Wasserstein-Fréchet regression and linear regression are shown in~\Cref{fig:accuracy_comparison}. As seen in~\Cref{fig:accuracy_comparison}, we can see that
the Wasserstein–Fréchet regression method outperforms the traditional linear regression method in terms of
accuracy. The accuracy of the Wasserstein–Fréchet regression method is higher than that of the traditional
linear regression method, indicating that the Wasserstein–Fréchet regression method is more effective in
capturing the underlying distribution of the data.
\section{Real Data Analysis}\label{sec:real_data_analysis}

In this section, we apply the proposed Wasserstein-Fréchet regression method to real-world neuroimaging data. The control dataset is obtained from the Cambridge Centre for Ageing and Neuroscience (Cam-CAN) data repository~\citep{cam-canCambridgeCentreAgeing2014}, containing approximately 700 subjects with T1-weighted structural MRI scans and 9-minute resting-state, eyes-closed magnetoencephalography (MEG) recordings. Subjects are aged between 18 and 87 years. The dataset is publicly available at~\href{https://camcan-archive.mrc-cbu.cam.ac.uk/dataaccess/}{Cam-CAN webpage}. The mTBI dataset is obtained from Innovision IP Ltd. This  dataset contains 143 subjects with T1-weighted structural MRI scans and two sessions of 5-minute resting-state, eyes-closed MEG recordings. Subjects are aged between 16 and 70 years. The dataset is not publicly available due to confidentiality agreements with patients. Both datasets were recorded using the MEGIN TRIUX system, a 306-channel MEG system with 102 magnetometers and 204 gradiometers. The goal of this analysis is to predict mTBI status based on demographic information (age, gender) and MRI/MEG data.

A promising MEG biomarker for mTBI detection is low-frequency power in resting-state recordings~\citep{allenMagnetoencephalographyAbnormalitiesAdult2021}. The prominent finding is that low-frequency power in resting-state MEG is significantly higher in mTBI patients than in healthy controls~\citep{lewineNeuromagneticAssessmentPathophysiologic1999,lewineObjectiveDocumentationTraumatic2007,huangSinglesubjectbasedWholebrainMEG2014,dunkley2015a}. Recent research~\citep{dunkley2015a,huangMEGSourceImaging2014,wangDisruptedGammaSynchrony2017,marshNetworkEffectsTraumatic2025} has also demonstrated statistically significant spectral power increases in the gamma band. To investigate the joint effects of delta and gamma bands, we focus on the following frequency bands of interest: delta (0.5--4 Hz)~\citep{hejaziSleepDeltaPower2024}, slow gamma (30--60 Hz), and fast gamma (60--100 Hz)~\citep{bieriSlowFastGamma2014}.

Next, we present a reproducible MRI+MEG analysis pipeline covering steps from raw data preprocessing to source magnitude imaging~\citep{gramfortMEGEEGData2013,grossGoodPracticeConducting2013,nisoBrainstormPipelineAnalysis2019,ferranteFLUXPipelineMEG2022}.

\textbf{Anatomical reconstruction.} We begin by constructing detailed brain models using FreeSurfer 8.0.0~\citep{fischlFreeSurfer2012}, an automated software suite for cortical surface reconstruction and volumetric segmentation. The T1-weighted structural MRI scans undergo the standard FreeSurfer pipeline `recon-all'. Following anatomical reconstruction, we construct Boundary Element Method (BEM) head models to obtain realistic forward models for MEG source analysis. BEM head model construction is performed using MNE-Python~\citep{gramfortMEGEEGData2013}.

\textbf{Data preprocessing.} The raw MEG data undergo comprehensive preprocessing using MNE-Python~\citep{gramfortMEGEEGData2013}. The preprocessing pipeline comprises several sequential steps: (1) Signal-Space Separation (SSS) is applied to eliminate environmental noise and artifacts from MEG recordings~\citep{tauluSpatiotemporalSignalSpace2006}; (2) Temporal cropping is performed to extract relevant time intervals, specifically from stimulus onset to termination when stimulus channels are present; (3) Power line artifact detection and removal is conducted to eliminate environmental interference manifesting as persistent oscillations at AC power line frequencies; (4) Temporal segmentation of continuous recordings into 2-second epochs is performed, excluding previously identified bad channels from analysis; (5) Computation of the noise covariance matrix is performed using approximately 5 minutes of empty room recordings to characterise environmental noise baseline for subsequent source modelling. After preprocessing, the control group (Cam-CAN) contains 588 healthy subjects, and the mTBI group contains 131 subjects.

\textbf{Source magnitude imaging.} Source magnitude imaging is performed using vector-based spatial-temporal analysis~\citep[VESTAL,][]{huangSinglesubjectbasedWholebrainMEG2014}. As the VESTAL algorithm is based on the \(L_1\)-minimum-norm solution, it is robust to environmental noise and artifacts in the data. MEG source magnitude imaging is carried out on preprocessed MEG data using the BEM head model. For each 2-second epoch, we apply band-pass filtering for the following frequency bands: delta (0.5--4 Hz), slow gamma (30--60 Hz), and fast gamma (60--100 Hz). For each frequency band, we obtain source magnitude images using the VESTAL algorithm~\citep{huangSinglesubjectbasedWholebrainMEG2014}. The brain volume is divided into a grid of 5120 nodes in the BEM model, meaning that the source magnitude image is a 5120-dimensional vector for each 2-second epoch. Rather than analyzing individual nodes, we focus on regions of interest (ROIs) defined by the Desikan-Killiany atlas~\citep{desikanAutomatedLabelingSystem2006}, which yields 68 regions across left and right brain hemispheres. For each region, we compute the average source magnitude across all nodes within the region. Finally, we concatenate the average source magnitude vectors across all epochs to form the final feature representation for each subject.

\textbf{Wasserstein-Fréchet regression.} First, training and test datasets are generated as follows: For the Cam-CAN dataset, we randomly select 131 subjects as the test dataset, with the remainder forming the training dataset. For the Innovision dataset, we randomly select 65 subjects as the test dataset, with the remaining 66 subjects forming the training dataset.

For each subject in the training datasets and each frequency band, we estimate quantile functions of magnitude using our diffusion-based kernel density estimation across all epochs for each brain area. We discretise the quantile function on the support \([0,1]\) with equal-grid spacing (1/1024), resulting in a finite number of quantile levels. Wasserstein-Fréchet regression from Section~\ref{sec:Methodology} is then performed on these quantile functions to model the relationship between MEG features and demographic information (age, gender).

In the second step, we generate predictions for each brain region and subject in our training data. We examine three frequency bands: delta, slow gamma, and fast gamma. The true labels are stored in vector \(Y_{0}\), where \(Y_{0,i}=1\) indicates mTBI and \(Y_{0,i}=0\) indicates control status for subject \(i\). Our prediction results create a binary matrix \(Z_0\) of size \((457+66, 68 \times 3)\).We consider using random forest to build two models: \(Y_0 \sim Z_0[:,1:68]\) and \(Y_0 \sim Z_0\), where \(Z_0[:,1:68]\) represents predicted status in the delta band.

Finally, we use the Wasserstein-Fréchet regression model to obtain a binary matrix \(Z_1\) with dimensions \((196, 204)\) for the test dataset. Let \(Y_1\) be the binary vector of underlying labels of test subjects. We use the trained random forest models to predict the status of each subject in the test dataset. Define \(\hat{Y}_{1,\text{delta}}\) and \(\hat{Y}_{1,\text{all}}\) as the predicted labels for delta-only and all-band analyses, respectively. We measure performance using standard classification accuracy \(ACC\). Results show that delta-band classification achieves 0.8893 accuracy. When incorporating all frequency bands, performance improves substantially to \textbf{0.9802}.

To provide a more comprehensive comparison with the linear and SVM classifiers mentioned in the literature, we trained and evaluated both classifiers on the same dataset used for our Wasserstein-Fréchet framework. First, since diffusion-based KDE can handle varying numbers of epochs across frequency bands and brain regions, while linear and SVM require balanced input data, we modified our feature extraction approach. Instead of using the full KDE representation, we computed the average spectral magnitude across all epochs for each region and frequency band as input features for the SVM and linear classifier.

Second, for the linear classifier baseline, we fitted the linear regression model from~\eqref{eq:linear_regression_model_sample} (described in~\Cref{subsec:WassersteinFréchet_regression}) to the balanced preprocessed data.
This approach generates binary predictions for individual brain regions and frequency bands. Following the same ensemble strategy as our Wasserstein-Fréchet approach, we employed random forest to aggregate individual predictions into final subject-level classifications. In essence, this baseline substitutes ordinary linear regression for the Wasserstein-Fréchet regression component while maintaining the same overall architecture.

Third, for the SVM classifier baseline, we constructed feature vectors using age, gender, and the averaged spectral magnitudes from each frequency band and brain region. For the delta-only analysis, this resulted in 70 features (68 brain regions plus age and gender). When incorporating all three frequency bands, the feature space expanded to 206 dimensions (68 regions \( \times \) 3 bands + 2 demographic variables). In our study, before implementation of SVM classification, we applied a standard scaler, i.e., standardised the features to have zero mean and unit variance, to the training and test data. We used the radial basis function (RBF) kernel for SVM classification, as it is effective in handling non-linear relationships in high-dimensional spaces~\citep{vapnikNatureStatisticalLearning1995,changLIBSVMLibrarySupport2011}. It is easy to implement the SVM classification in Python using libraries such as~\href{https://scikit-learn.org/stable/}{scikit-learn}. Finally, when analyzing only the delta band, the linear and SVM classifiers achieved test accuracies of 0.8776 and 0.9031, respectively. For the analysis incorporating all three frequency bands, the linear and SVM classifiers obtained test accuracies of 0.9388 and 0.9337, respectively.
Both accuracies are lower than 0.9802 achieved by our Wasserstein-Fréchet regression approach. This is not surprising because the linear and SVM classifiers use only the first moment of distributions instead of the full density. The linear and SVM classifiers discards higher-order moments and distributional shape information that may be important for classification.
\begin{table}[ht]
    \centering
    \caption{Prediction accuracy comparison for mild traumatic brain injury (mTBI) detection. Baselines 1 and 2 correspond to the linear and SVM classifiers based on average spectral magnitudes, respectively. We only report the accuracies for all frequency bands.}\label{tab:mTBI_accuracy_comparison}
    \resizebox{0.95\textwidth}{!}{
        \begin{tabular}{lccc}
            \toprule
            References & Accuracy & Based-on & Method \\
            \midrule
            \citet{huangSinglesubjectbasedWholebrainMEG2014} & 0.83 & Resting-state MEG, Age, Gender & Statistical Model (VESTAL) \\
            \citet{vakorinDetectingMildTraumatic2016} & 0.88 & Resting-state MEG, Connectivity & Machine Learning \\
            \multirow{2}{*}{\citet{vergaraDetectionMildTraumatic2017}} & \multirow{2}{*}{0.841} & Resting-state MEG, Connectivity & \multirow{2}{*}{Support Vector Machine (SVM)} \\
            &  & and Fractional Anisotropy & \\
            \citet{mcnerneyObjectiveClassificationMTBI2019} & 0.91 & EEG+Self-reported Symptoms & Machine Learning \\
            \citet{thanjavurRecurrentNeuralNetworkbased2021} & 0.926 & EEG & Recurrent Neural Network (RNN) \\
            \citet{italinnaUsingNormativeModeling2023} & 0.79 & Resting-state MEG, Source Magnitude & Support Vector Machine (SVM) \\
            Baseline 1 & 0.9388 & Resting-state MEG, Source Magnitude & OLS Classifiers \\
            Baseline 2 & 0.9337 & Resting-state MEG, Source Magnitude & SVM Classifiers \\
            Our method & \textbf{0.9802} & Resting-state MEG, Source Magnitude & Wasserstein-Fréchet Regression \\
            \bottomrule
        \end{tabular}
    }
\end{table}

From the~\Cref{tab:mTBI_accuracy_comparison}, we can see that our method achieves the highest prediction accuracy of 0.9802, which is significantly higher than the previous methods. This indicates that the Wasserstein-Fréchet regression method is effective in capturing the underlying distribution of the source magnitude across epochs and can be used as a powerful tool for mTBI detection.


\section{Conclusion}\label{sec:conclusion}
This study introduced a quantile function-based representation of epoch-wise MEG source magnitude distributions and a global Wasserstein-Fréchet regression framework to link these functional responses to demographic covariates. Methodologically, the work contributes (i) a diffusion-based kernel density procedure with fixed-point bandwidth selection and fast spectral approximation to estimate subject- and region-specific quantile functions; (ii) a covariate-adjusted Wasserstein regression that yields conditional mean distributions without local smoothing or tuning; and (iii) a two-stage classification strategy that aggregates region-level predictions across three frequency bands. The pipeline relies on anatomically realistic forward modelling using BEM, and utilises \( L_1 \)-based source magnitude imaging (VESTAL) to stabilise estimates under realistic noise conditions.

Simulation results demonstrate that the proposed density estimation reduces total variation error relative to a recent diffusion-based KDE alternative across sample sizes. Wasserstein-Fréchet regression outperforms standard linear regression when the response is a fundamental quantile function. On real data combining Cam-CAN controls and  mTBI cases, the approach attains high predictive performance, with accuracy of 0.9802 when integrating delta, slow gamma, and fast gamma bands, exceeding delta-only performance. Our result also align with reports of low-frequency and gamma-band abnormalities in mTBI.\ These gains reflect the joint benefits of (a) preserving full distributional information via quantile functions, (b) modelling covariate effects at the distribution level, (c) multiband aggregation that mitigates frequency-specific leakage and enhances robustness.

Limitations include the need for broader external validation beyond a single  cohort, assessment of generalisability across sites, scanners, and preprocessing choices (e.g., SSS parameters, epoching, ROI definitions). Future work will pursue multicenter validation, joint modelling of power and connectivity. Overall, the results support Wasserstein-Fréchet regression on quantile representations as a principled and effective framework for MEG biomarkers in mTBI detection, with clear translational potential given its accuracy, modularity, and compatibility with standard neuroimaging.
\section*{Data and Code Availability}
The control data comes from the \href{https://cam-can.mrc-cbu.cam.ac.uk/dataset/}{Cambridge Centre for Ageing and Neuroscience (Cam-CAN) dataset}~\citep{cam-canCambridgeCentreAgeing2014}. The code for the simulation studies are available in the Python package \href{https://github.com/innovision-ip/NeuroWAR}{NeuroWAR}.

\section*{Competing interests}
The authors from the University of Kent, J.L., J.Z.. have no competing interests.\ G.G. is an employee of Innovision IP Ltd, which provides commercial reports on individuals who may have had a head injury.

\section*{Funding}
This research received financial support jointly from UKRI through Innovate UK and Innovision IP Ltd. (Grant No: 10053865).
\section*{Acknowledgments}
This work was supported by the High-Performance Computing Cluster at the University of Kent.
\bibliographystyle{chicago}
\bibliography{ref-wd.bib}

\begin{thebibliography}{}

\bibitem[\protect\citeauthoryear{Allen, Halsey, Topcu, Rier, Gascoyne, Scadding, Furlong, Dunkley, {das Nair}, Brookes, and Evangelou}{Allen et~al.}{2021}]{allenMagnetoencephalographyAbnormalitiesAdult2021}
Allen, C.~M., L.~Halsey, G.~Topcu, L.~Rier, L.~E. Gascoyne, J.~W. Scadding, P.~L. Furlong, B.~T. Dunkley, R.~{das Nair}, M.~J. Brookes, and N.~Evangelou (2021).
\newblock Magnetoencephalography abnormalities in adult mild traumatic brain injury: {{A}} systematic review.
\newblock {\em NeuroImage : Clinical\/}~{\em 31}, 102697.

\bibitem[\protect\citeauthoryear{Antonakakis, Dimitriadis, Zervakis, Papanicolaou, and Zouridakis}{Antonakakis et~al.}{2017}]{antonakakisAlteredRichClubFrequencyDependent2017}
Antonakakis, M., S.~I. Dimitriadis, M.~Zervakis, A.~C. Papanicolaou, and G.~Zouridakis (2017).
\newblock Altered {{Rich-Club}} and {{Frequency-Dependent Subnetwork Organization}} in {{Mild Traumatic Brain Injury}}: {{A MEG Resting-State Study}}.
\newblock {\em Frontiers in Human Neuroscience\/}~{\em 11}, 416.

\bibitem[\protect\citeauthoryear{Barredo~Arrieta, {D{\'i}az-Rodr{\'i}guez}, Del~Ser, Bennetot, Tabik, Barbado, Garcia, {Gil-Lopez}, Molina, Benjamins, Chatila, and Herrera}{Barredo~Arrieta et~al.}{2020}]{barredoarrietaExplainableArtificialIntelligence2020}
Barredo~Arrieta, A., N.~{D{\'i}az-Rodr{\'i}guez}, J.~Del~Ser, A.~Bennetot, S.~Tabik, A.~Barbado, S.~Garcia, S.~{Gil-Lopez}, D.~Molina, R.~Benjamins, R.~Chatila, and F.~Herrera (2020).
\newblock Explainable {{Artificial Intelligence}} ({{XAI}}): {{Concepts}}, taxonomies, opportunities and challenges toward responsible {{AI}}.
\newblock {\em Information Fusion\/}~{\em 58}, 82--115.

\bibitem[\protect\citeauthoryear{Bieri, Bobbitt, and Colgin}{Bieri et~al.}{2014}]{bieriSlowFastGamma2014}
Bieri, K.~W., K.~N. Bobbitt, and L.~L. Colgin (2014).
\newblock Slow and fast gamma rhythms coordinate different spatial coding modes in hippocampal place cells.
\newblock {\em Neuron\/}~{\em 82\/}(3), 670--681.

\bibitem[\protect\citeauthoryear{Bollaerts, Eilers, and Aerts}{Bollaerts et~al.}{2006}]{bollaertsQuantileRegressionMonotonicity2006}
Bollaerts, K., P.~H. Eilers, and M.~Aerts (2006).
\newblock Quantile regression with monotonicity restrictions using {{P-splines}} and the {{L1-norm}}.
\newblock {\em Statistical Modelling\/}~{\em 6\/}(3), 189--207.

\bibitem[\protect\citeauthoryear{Boon, Tewarie, Berendse, Stam, and Hillebrand}{Boon et~al.}{2021}]{boonLongitudinalConsistencySourcespace2021}
Boon, L.~I., P.~Tewarie, H.~W. Berendse, C.~J. Stam, and A.~Hillebrand (2021).
\newblock Longitudinal consistency of source-space spectral power and functional connectivity using different magnetoencephalography recording systems.
\newblock {\em Scientific Reports\/}~{\em 11\/}(1), 16336.

\bibitem[\protect\citeauthoryear{Botev, Grotowski, and Kroese}{Botev et~al.}{2010}]{botevKernelDensityEstimation2010}
Botev, Z.~I., J.~F. Grotowski, and D.~P. Kroese (2010).
\newblock Kernel density estimation via diffusion.
\newblock {\em The Annals of Statistics\/}~{\em 38\/}(5), 2916--2957.

\bibitem[\protect\citeauthoryear{Bowman, Hall, and Prvan}{Bowman et~al.}{1998}]{bowmanBandwidthSelectionSmoothing1998}
Bowman, A., P.~Hall, and T.~Prvan (1998).
\newblock Bandwidth {{Selection}} for the {{Smoothing}} of {{Distribution Functions}}.
\newblock {\em Biometrika\/}~{\em 85\/}(4), 799--808.

\bibitem[\protect\citeauthoryear{Breiman}{Breiman}{2001}]{breimanStatisticalModelingTwo2001}
Breiman, L. (2001).
\newblock Statistical {{Modeling}}: {{The Two Cultures}} (with comments and a rejoinder by the author).
\newblock {\em Statistical Science\/}~{\em 16\/}(3), 199--231.

\bibitem[\protect\citeauthoryear{Caiola, Babu, and Ye}{Caiola et~al.}{2023}]{caiolaEEGClassificationTraumatic2023}
Caiola, M., A.~Babu, and M.~Ye (2023).
\newblock {{EEG}} classification of traumatic brain injury and stroke from a nonspecific population using neural networks.
\newblock {\em PLOS Digital Health\/}~{\em 2\/}(7), e0000282.

\bibitem[\protect\citeauthoryear{{Cam-CAN}, Shafto, Tyler, Dixon, Taylor, Rowe, Cusack, Calder, {Marslen-Wilson}, Duncan, Dalgleish, Henson, Brayne, and Matthews}{{Cam-CAN} et~al.}{2014}]{cam-canCambridgeCentreAgeing2014}
{Cam-CAN}, M.~A. Shafto, L.~K. Tyler, M.~Dixon, J.~R. Taylor, J.~B. Rowe, R.~Cusack, A.~J. Calder, W.~D. {Marslen-Wilson}, J.~Duncan, T.~Dalgleish, R.~N. Henson, C.~Brayne, and F.~E. Matthews (2014).
\newblock The {{Cambridge Centre}} for {{Ageing}} and {{Neuroscience}} ({{Cam-CAN}}) study protocol: A cross-sectional, lifespan, multidisciplinary examination of healthy cognitive ageing.
\newblock {\em BMC Neurology\/}~{\em 14\/}(1), 204.

\bibitem[\protect\citeauthoryear{Chang and Lin}{Chang and Lin}{2011}]{changLIBSVMLibrarySupport2011}
Chang, C.-C. and C.-J. Lin (2011).
\newblock {{LIBSVM}}: {{A}} library for support vector machines.
\newblock {\em ACM Transactions on Intelligent Systems and Technology\/}~{\em 2\/}(3), 1--27.

\bibitem[\protect\citeauthoryear{Desikan, S{\'e}gonne, Fischl, Quinn, Dickerson, Blacker, Buckner, Dale, Maguire, Hyman, Albert, and Killiany}{Desikan et~al.}{2006}]{desikanAutomatedLabelingSystem2006}
Desikan, R.~S., F.~S{\'e}gonne, B.~Fischl, B.~T. Quinn, B.~C. Dickerson, D.~Blacker, R.~L. Buckner, A.~M. Dale, R.~P. Maguire, B.~T. Hyman, M.~S. Albert, and R.~J. Killiany (2006).
\newblock An automated labeling system for subdividing the human cerebral cortex on {{MRI}} scans into gyral based regions of interest.
\newblock {\em NeuroImage\/}~{\em 31\/}(3), 968--980.

\bibitem[\protect\citeauthoryear{Dunkley, Costa, Bethune, Jetly, Pang, Taylor, and Doesburg}{Dunkley et~al.}{2015}]{dunkley2015a}
Dunkley, B., L.~Costa, A.~Bethune, R.~Jetly, E.~Pang, M.~Taylor, and S.~Doesburg (2015).
\newblock Low-frequency connectivity is associated with mild traumatic brain injury.
\newblock {\em NeuroImage: Clinical\/}~{\em 7}, 611--621.

\bibitem[\protect\citeauthoryear{Ferrante, Liu, Minarik, Gorska, Ghafari, Luo, and Jensen}{Ferrante et~al.}{2022}]{ferranteFLUXPipelineMEG2022}
Ferrante, O., L.~Liu, T.~Minarik, U.~Gorska, T.~Ghafari, H.~Luo, and O.~Jensen (2022).
\newblock {{FLUX}}: {{A}} pipeline for {{MEG}} analysis.
\newblock {\em NeuroImage\/}~{\em 253}, 119047.

\bibitem[\protect\citeauthoryear{Fischl}{Fischl}{2012}]{fischlFreeSurfer2012}
Fischl, B. (2012).
\newblock {{FreeSurfer}}.
\newblock {\em NeuroImage\/}~{\em 62\/}(2), 774--781.

\bibitem[\protect\citeauthoryear{Gramfort}{Gramfort}{2013}]{gramfortMEGEEGData2013}
Gramfort, A. (2013).
\newblock {{MEG}} and {{EEG}} data analysis with {{MNE-Python}}.
\newblock {\em Frontiers in Neuroscience\/}~{\em 7}, 1--13.

\bibitem[\protect\citeauthoryear{Gross, Baillet, Barnes, Henson, Hillebrand, Jensen, Jerbi, Litvak, Maess, Oostenveld, Parkkonen, Taylor, Van~Wassenhove, Wibral, and Schoffelen}{Gross et~al.}{2013}]{grossGoodPracticeConducting2013}
Gross, J., S.~Baillet, G.~R. Barnes, R.~N. Henson, A.~Hillebrand, O.~Jensen, K.~Jerbi, V.~Litvak, B.~Maess, R.~Oostenveld, L.~Parkkonen, J.~R. Taylor, V.~Van~Wassenhove, M.~Wibral, and J.-M. Schoffelen (2013).
\newblock Good practice for conducting and reporting {{MEG}} research.
\newblock {\em NeuroImage\/}~{\em 65}, 349--363.

\bibitem[\protect\citeauthoryear{Hejazi, Duncan, Kheirkhah, Kowalczyk, Riedner, Oppenheimer, Momenan, Yuan, Kerich, Goldman, and Zarate}{Hejazi et~al.}{2024}]{hejaziSleepDeltaPower2024}
Hejazi, N.~S., W.~C. Duncan, M.~Kheirkhah, A.~Kowalczyk, B.~Riedner, M.~Oppenheimer, R.~Momenan, Q.~Yuan, M.~Kerich, D.~Goldman, and C.~A. Zarate (2024).
\newblock Sleep {{Delta}} power, age, and sex effects in treatment-resistant depression.
\newblock {\em Journal of Psychiatric Research\/}~{\em 174}, 332--339.

\bibitem[\protect\citeauthoryear{Huang, Huang, Robb, Angeles, Nichols, Baker, Song, Harrington, Theilmann, Srinivasan, Heister, Diwakar, Canive, Edgar, Chen, Ji, Shen, {El-Gabalawy}, Levy, McLay, {Webb-Murphy}, Liu, Drake, and Lee}{Huang et~al.}{2014}]{huangMEGSourceImaging2014}
Huang, M.-X., C.~W. Huang, A.~Robb, A.~Angeles, S.~L. Nichols, D.~G. Baker, T.~Song, D.~L. Harrington, R.~J. Theilmann, R.~Srinivasan, D.~Heister, M.~Diwakar, J.~M. Canive, J.~C. Edgar, Y.-H. Chen, Z.~Ji, M.~Shen, F.~{El-Gabalawy}, M.~Levy, R.~McLay, J.~{Webb-Murphy}, T.~T. Liu, A.~Drake, and R.~R. Lee (2014).
\newblock {{MEG}} source imaging method using fast {{L1}} minimum-norm and its applications to signals with brain noise and human resting-state source amplitude images.
\newblock {\em NeuroImage\/}~{\em 84}, 585--604.

\bibitem[\protect\citeauthoryear{Huang, Nichols, Baker, Robb, Angeles, Yurgil, Drake, Levy, Song, McLay, Theilmann, Diwakar, Risbrough, Ji, Huang, Chang, Harrington, Muzzatti, Canive, Christopher~Edgar, Chen, and Lee}{Huang et~al.}{2014}]{huangSinglesubjectbasedWholebrainMEG2014}
Huang, M.-X., S.~Nichols, D.~G. Baker, A.~Robb, A.~Angeles, K.~A. Yurgil, A.~Drake, M.~Levy, T.~Song, R.~McLay, R.~J. Theilmann, M.~Diwakar, V.~B. Risbrough, Z.~Ji, C.~W. Huang, D.~G. Chang, D.~L. Harrington, L.~Muzzatti, J.~M. Canive, J.~Christopher~Edgar, Y.-H. Chen, and R.~R. Lee (2014).
\newblock Single-subject-based whole-brain {{MEG}} slow-wave imaging approach for detecting abnormality in patients with mild traumatic brain injury.
\newblock {\em NeuroImage : Clinical\/}~{\em 5}, 109--119.

\bibitem[\protect\citeauthoryear{It{\"a}linna, Kaltiainen, Forss, Liljestr{\"o}m, and Parkkonen}{It{\"a}linna et~al.}{2023}]{italinnaUsingNormativeModeling2023}
It{\"a}linna, V., H.~Kaltiainen, N.~Forss, M.~Liljestr{\"o}m, and L.~Parkkonen (2023).
\newblock Using normative modeling and machine learning for detecting mild traumatic brain injury from magnetoencephalography data.
\newblock {\em PLOS Computational Biology\/}~{\em 19\/}(11), e1011613.

\bibitem[\protect\citeauthoryear{{Law Commission of England and Wales}}{{Law Commission of England and Wales}}{2011}]{lawcommissionExpertEvidence2011}
{Law Commission of England and Wales} (2011).
\newblock Expert evidence in criminal proceedings in england and wales.
\newblock Law Com No 325.

\bibitem[\protect\citeauthoryear{Leslie}{Leslie}{2020}]{leslieExplainingDecisionsMade2020}
Leslie, D. (2020).
\newblock Explaining {{Decisions Made}} with {{AI}}.

\bibitem[\protect\citeauthoryear{Lewine, Davis, Bigler, Thoma, Hill, Funke, Sloan, Hall, and Orrison}{Lewine et~al.}{2007}]{lewineObjectiveDocumentationTraumatic2007}
Lewine, J.~D., J.~T. Davis, E.~D. Bigler, R.~Thoma, D.~Hill, M.~Funke, J.~H. Sloan, S.~Hall, and W.~W. Orrison (2007).
\newblock Objective {{Documentation}} of {{Traumatic Brain Injury Subsequent}} to {{Mild Head Trauma}}: {{Multimodal Brain Imaging With MEG}}, {{SPECT}}, and {{MRI}}.
\newblock {\em The Journal of Head Trauma Rehabilitation\/}~{\em 22\/}(3), 141.

\bibitem[\protect\citeauthoryear{Lewine, Davis, Sloan, Kodituwakku, and Jr}{Lewine et~al.}{1999}]{lewineNeuromagneticAssessmentPathophysiologic1999}
Lewine, J.~D., J.~T. Davis, J.~H. Sloan, P.~W. Kodituwakku, and W.~W.~O. Jr (1999).
\newblock Neuromagnetic {{Assessment}} of {{Pathophysiologic Brain Activity Induced}} by {{Minor Head Trauma}}.
\newblock {\em American Journal of Neuroradiology\/}~{\em 20\/}(5), 857--866.

\bibitem[\protect\citeauthoryear{Li, Green, Carr, Liu, and Zhang}{Li et~al.}{2025}]{liBayesianInferenceGeneral2025}
Li, J., G.~Green, S.~J.~A. Carr, P.~Liu, and J.~Zhang (2025).
\newblock Bayesian {{Inference General Procedures}} for {{A Single-subject Test}} study.
\newblock {\em Neuroscience Informatics\/}~{\em 5\/}(2), 100195.

\bibitem[\protect\citeauthoryear{Marron and Ruppert}{Marron and Ruppert}{1994}]{marronTransformationsReduceBoundary1994}
Marron, J.~S. and D.~Ruppert (1994).
\newblock Transformations to {{Reduce Boundary Bias}} in {{Kernel Density Estimation}}.
\newblock {\em Journal of the Royal Statistical Society: Series B (Methodological)\/}~{\em 56\/}(4), 653--671.

\bibitem[\protect\citeauthoryear{Marsh, Chauvette, Huang, Timofeev, and Bazhenov}{Marsh et~al.}{2025}]{marshNetworkEffectsTraumatic2025}
Marsh, B., S.~Chauvette, M.~Huang, I.~Timofeev, and M.~Bazhenov (2025).
\newblock Network effects of traumatic brain injury: From infra slow to high frequency oscillations and seizures.
\newblock {\em Journal of Computational Neuroscience\/}~{\em 53\/}(2), 247--266.

\bibitem[\protect\citeauthoryear{McNerney, Hobday, Cole, Ganong, Winans, Matthews, Hood, and Lane}{McNerney et~al.}{2019}]{mcnerneyObjectiveClassificationMTBI2019}
McNerney, M.~W., T.~Hobday, B.~Cole, R.~Ganong, N.~Winans, D.~Matthews, J.~Hood, and S.~Lane (2019).
\newblock Objective {{Classification}} of {{mTBI Using Machine Learning}} on a {{Combination}} of {{Frontopolar Electroencephalography Measurements}} and {{Self-reported Symptoms}}.
\newblock {\em Sports Medicine - Open\/}~{\em 5}, 14.

\bibitem[\protect\citeauthoryear{Niso, Tadel, Bock, Cousineau, Santos, and Baillet}{Niso et~al.}{2019}]{nisoBrainstormPipelineAnalysis2019}
Niso, G., F.~Tadel, E.~Bock, M.~Cousineau, A.~Santos, and S.~Baillet (2019).
\newblock Brainstorm {{Pipeline Analysis}} of {{Resting-State Data From}} the {{Open MEG Archive}}.
\newblock {\em Frontiers in Neuroscience\/}~{\em 13}, 1--10.

\bibitem[\protect\citeauthoryear{Park, ~, ~, and {and Kang}}{Park et~al.}{2003}]{parkAdaptiveVariableLocation2003}
Park, B.~U., J.~, Seok-Oh, J.~, M.~C., and K.-H. {and Kang} (2003).
\newblock Adaptive variable location kernel density estimators with good performance at boundaries.
\newblock {\em Journal of Nonparametric Statistics\/}~{\em 15\/}(1), 61--75.

\bibitem[\protect\citeauthoryear{Parzen}{Parzen}{1979}]{parzenNonparametricStatisticalData1979}
Parzen, E. (1979).
\newblock Nonparametric {{Statistical Data Modeling}}.
\newblock {\em Journal of the American Statistical Association\/}~{\em 74\/}(365), 105--121.

\bibitem[\protect\citeauthoryear{Pelz, Schartau, Somes, Lampe, and Slawig}{Pelz et~al.}{2023}]{pelzDiffusionbasedKernelDensity2023}
Pelz, M.-T., M.~Schartau, C.~J. Somes, V.~Lampe, and T.~Slawig (2023).
\newblock A diffusion-based kernel density estimator ({{diffKDE}}, version 1) with optimal bandwidth approximation for the analysis of data in geoscience and ecological research.
\newblock {\em Geoscientific Model Development\/}~{\em 16\/}(22), 6609--6634.

\bibitem[\protect\citeauthoryear{Petersen, Liu, and Divani}{Petersen et~al.}{2021}]{petersenWasserstein$F$testsConfidence2021}
Petersen, A., X.~Liu, and A.~A. Divani (2021).
\newblock Wasserstein \${{F}}\$-tests and confidence bands for the {{Fr{\'e}chet}} regression of density response curves.
\newblock {\em The Annals of Statistics\/}~{\em 49\/}(1), 590--611.

\bibitem[\protect\citeauthoryear{Petersen and M{\"u}ller}{Petersen and M{\"u}ller}{2016}]{petersenFunctionalDataAnalysis2016}
Petersen, A. and H.-G. M{\"u}ller (2016).
\newblock Functional {{Data Analysis}} for {{Density Functions}} by {{Transformation}} to a {{Hilbert Space}}.
\newblock {\em The Annals of Statistics\/}~{\em 44\/}(1), 183--218.

\bibitem[\protect\citeauthoryear{Petersen and M{\"u}ller}{Petersen and M{\"u}ller}{2019}]{petersenFrechetRegressionRandom2019}
Petersen, A. and H.-G. M{\"u}ller (2019).
\newblock Fr{\'e}chet regression for random objects with {{Euclidean}} predictors.
\newblock {\em The Annals of Statistics\/}~{\em 47\/}(2), 691--719.

\bibitem[\protect\citeauthoryear{Petersen, Zhang, and Kokoszka}{Petersen et~al.}{2022}]{petersenModelingProbabilityDensity2022}
Petersen, A., C.~Zhang, and P.~Kokoszka (2022).
\newblock Modeling {{Probability Density Functions}} as {{Data Objects}}.
\newblock {\em Econometrics and Statistics\/}~{\em 21}, 159--178.

\bibitem[\protect\citeauthoryear{Ramsay and Silverman}{Ramsay and Silverman}{2006}]{ramsayFunctionalDataAnalysis2006}
Ramsay, J.~O. and B.~W. Silverman (2006).
\newblock {\em Functional Data Analysis\/} (2. ed., [Nachdr.] ed.).
\newblock Springer Series in Statistics. New York, NY: Springer.

\bibitem[\protect\citeauthoryear{Rudin}{Rudin}{2019}]{rudinStopExplainingBlack2019}
Rudin, C. (2019).
\newblock Stop explaining black box machine learning models for high stakes decisions and use interpretable models instead.
\newblock {\em Nature Machine Intelligence\/}~{\em 1\/}(5), 206--215.

\bibitem[\protect\citeauthoryear{Takeuchi, Le, Sears, and Smola}{Takeuchi et~al.}{2006}]{takeuchiNonparametricQuantileEstimation2006}
Takeuchi, I., Q.~V. Le, T.~D. Sears, and A.~J. Smola (2006).
\newblock Nonparametric {{Quantile Estimation}}.
\newblock {\em Journal of machine learning research\/}~{\em 7\/}(7), 1231--1264.

\bibitem[\protect\citeauthoryear{Taulu and Simola}{Taulu and Simola}{2006}]{tauluSpatiotemporalSignalSpace2006}
Taulu, S. and J.~Simola (2006).
\newblock Spatiotemporal signal space separation method for rejecting nearby interference in {{MEG}} measurements.
\newblock {\em Physics in Medicine and Biology\/}~{\em 51\/}(7), 1759--1768.

\bibitem[\protect\citeauthoryear{Terrell and Scott}{Terrell and Scott}{1992}]{terrellVariableKernelDensity1992}
Terrell, G.~R. and D.~W. Scott (1992).
\newblock Variable {{Kernel Density Estimation}}.
\newblock {\em The Annals of Statistics\/}~{\em 20\/}(3), 1236--1265.

\bibitem[\protect\citeauthoryear{Thanjavur, Babul, Foran, Bielecki, Gilchrist, Hristopulos, Brucar, and {Virji-Babul}}{Thanjavur et~al.}{2021}]{thanjavurRecurrentNeuralNetworkbased2021}
Thanjavur, K., A.~Babul, B.~Foran, M.~Bielecki, A.~Gilchrist, D.~T. Hristopulos, L.~R. Brucar, and N.~{Virji-Babul} (2021).
\newblock Recurrent neural network-based acute concussion classifier using raw resting state {{EEG}} data.
\newblock {\em Scientific Reports\/}~{\em 11\/}(1), 12353.

\bibitem[\protect\citeauthoryear{Vakorin, Doesburg, da~Costa, Jetly, Pang, and Taylor}{Vakorin et~al.}{2016}]{vakorinDetectingMildTraumatic2016}
Vakorin, V.~A., S.~M. Doesburg, L.~da~Costa, R.~Jetly, E.~W. Pang, and M.~J. Taylor (2016).
\newblock Detecting {{Mild Traumatic Brain Injury Using Resting State Magnetoencephalographic Connectivity}}.
\newblock {\em PLOS Computational Biology\/}~{\em 12\/}(12), e1004914.

\bibitem[\protect\citeauthoryear{Vapnik}{Vapnik}{1995}]{vapnikNatureStatisticalLearning1995}
Vapnik, V.~N. (1995).
\newblock {\em The {{Nature}} of {{Statistical Learning Theory}}}.
\newblock New York, NY: Springer New York.

\bibitem[\protect\citeauthoryear{Vergara, Mayer, Damaraju, Kiehl, and Calhoun}{Vergara et~al.}{2017}]{vergaraDetectionMildTraumatic2017}
Vergara, V.~M., A.~R. Mayer, E.~Damaraju, K.~A. Kiehl, and V.~Calhoun (2017).
\newblock Detection of {{Mild Traumatic Brain Injury}} by {{Machine Learning Classification Using Resting State Functional Network Connectivity}} and {{Fractional Anisotropy}}.
\newblock {\em Journal of Neurotrauma\/}~{\em 34\/}(5), 1045--1053.

\bibitem[\protect\citeauthoryear{Vivaldi, Caiola, Solarana, and Ye}{Vivaldi et~al.}{2021}]{vivaldiEvaluatingPerformanceEEG2021}
Vivaldi, N., M.~Caiola, K.~Solarana, and M.~Ye (2021).
\newblock Evaluating {{Performance}} of {{EEG Data-Driven Machine Learning}} for {{Traumatic Brain Injury Classification}}.
\newblock {\em IEEE transactions on bio-medical engineering\/}~{\em 68\/}(11), 3205--3216.

\bibitem[\protect\citeauthoryear{Wall, Powell, Young, Zynda, Stuart, Covassin, and Godfrey}{Wall et~al.}{2022}]{wallDeepLearningbasedApproach2022}
Wall, C., D.~Powell, F.~Young, A.~J. Zynda, S.~Stuart, T.~Covassin, and A.~Godfrey (2022).
\newblock A deep learning-based approach to diagnose mild traumatic brain injury using audio classification.
\newblock {\em PLOS ONE\/}~{\em 17\/}(9), e0274395.

\bibitem[\protect\citeauthoryear{Wand and Jones}{Wand and Jones}{1995}]{wandKernelSmoothing1995}
Wand, M.~P. and M.~C. Jones (1995).
\newblock {\em Kernel Smoothing\/} (1st ed ed.).
\newblock Number~60 in Monographs on Statistics and Applied Probability. London ; New York: Chapman \& Hall.

\bibitem[\protect\citeauthoryear{Wang, Costanzo, Rapp, Darmon, Nathan, Bashirelahi, Pham, Roy, and Keyser}{Wang et~al.}{2017}]{wangDisruptedGammaSynchrony2017}
Wang, C., M.~E. Costanzo, P.~E. Rapp, D.~Darmon, D.~E. Nathan, K.~Bashirelahi, D.~L. Pham, M.~J. Roy, and D.~O. Keyser (2017).
\newblock Disrupted {{Gamma Synchrony}} after {{Mild Traumatic Brain Injury}} and {{Its Correlation}} with {{White Matter Abnormality}}.
\newblock {\em Frontiers in Neurology\/}~{\em 8}, 571.

\bibitem[\protect\citeauthoryear{Wang, Chiou, and M{\"u}ller}{Wang et~al.}{2016}]{wangFunctionalDataAnalysis2016}
Wang, J.-L., J.-M. Chiou, and H.-G. M{\"u}ller (2016).
\newblock Functional {{Data Analysis}}.
\newblock {\em Annual Review of Statistics and Its Application\/}~{\em 3\/}(1), 257--295.

\bibitem[\protect\citeauthoryear{{World Health Organization}}{{World Health Organization}}{2021}]{WHOAIHealth2021}
{World Health Organization} (2021).
\newblock Ethics and governance of artificial intelligence for health.
\newblock World Health Organization.

\end{thebibliography}
\end{document}